\documentclass[final,twocolumn,showpacs,superscriptaddress,floatfix]{revtex4}
\usepackage{graphicx}

\usepackage{bm}
\usepackage{amsmath}
\usepackage{amsbsy}
\usepackage{amssymb}
\usepackage{amsfonts}

\renewcommand{\vec}[1]{{\bm #1}} 

 \newcommand{\pdiff}[3][]{\frac{\partial^{#1} #2}{\partial #3^{#1}}}     
                    
\newcommand{\dens}[2]{#1\times 10^{#2}\,\text{cm}^{-2}}

\begin{document}

\title{Influence of carrier-carrier and carrier-phonon correlations 
on optical absorption and gain in quantum-dot systems}

\author{M. Lorke}
\email{www.itp.uni-bremen.de/ag-jahnke/}
\affiliation{Institute for Theoretical Physics,  
             University of Bremen,
             28334 Bremen, Germany}

\author{T. R. Nielsen} 
\affiliation{Institute for Theoretical Physics,  
             University of Bremen,
             28334 Bremen, Germany}

\author{J. Seebeck}
\affiliation{Institute for Theoretical Physics,  
             University of Bremen,
             28334 Bremen, Germany}
	 
\author{P. Gartner}
\affiliation{Institute for Theoretical Physics,  
             University of Bremen,
             28334 Bremen, Germany}
\affiliation{National Institute for Materials Physics, POB MG-7, 
             Bucharest-Magurele, Romania}

\author{F. Jahnke}
\affiliation{Institute for Theoretical Physics,  
             University of Bremen,
             28334 Bremen, Germany}

\date{\today}

\pacs{78.67.Hc, 71.35.Cc}

\begin{abstract}
A microscopic theory is used to study the optical properties of semiconductor quantum dots.
The dephasing of a coherent excitation and line-shifts of the interband transitions 
due to carrier-carrier Coulomb interaction and carrier-phonon
interaction are determined from a quantum kinetic treatment of correlation processes. 
We investigate the density dependence of both mechanisms
and clarify the importance of various dephasing channels involving the localized 
and delocalized states of the system.

\end{abstract}

\maketitle

\section{Introduction}

In recent years, semiconductor quantum dots (QDs) have been studied extensively due to
possible applications in optoelectronic devices like LEDs, lasers, or amplifiers
\cite{Masumoto:02,Michler:03}. In the rapidly emerging field of quantum information technology, 
QDs have been successfully used
to demonstrate the generation of single photons or correlated photon pairs \cite{Michler:00,Moreau:01,Pelton:02}.
Furthermore, the strong coupling regime for QD emitters in optical microcavities has been demonstrated 
\cite{Reithmaier:04,Yoshie:04}. A common aspect in fundamental studies 
and practical applications of QDs is the critical
role of dephasing processes. They determine the homogeneous linewidth of the QD resonances, limit the coherence 
properties of QD lasers and their ultrafast emission dynamics, and have a strong influence on coherent
optical nonlinearities. Moreover, dephasing processes are intimately linked to lineshifts of the QD resonances.

Optical studies of QDs have been recently focused on self-assembled systems which are typically grown in the 
Stranski-Krastanoff mode. The resulting QDs are randomly distributed on a two-dimensional wetting layer (WL).
The interplay of carriers in the localized QD and delocalized WL states is mediated by 
the Coulomb interaction of carriers in addition to carrier-phonon interaction.

Theoretical studies of absorption and gain in QDs have been accomplished on the basis of
multilevel optical Bloch equations \cite{Hu:96}, or semiconductor Bloch equations (SBE) with
screened exchange and Coulomb hole contributions \cite{Schneider:01a}. 
However, in these approaches the dephasing is merely a parameter entering the calculation 
via a $T_2$-time. In Ref.~\cite{Schneider:04} the dephasing due
to Coulomb interaction has been calculated using SBE with correlation contributions 
due to Auger-like WL-assisted capture processes (see below) where scattering integrals have been evaluated in 
terms of free-carrier energies. A more elaborate analysis of dephasing due to Coulomb interaction in 
quantum wells (QWs),
which included non-Markovian scattering integrals based on renormalized energies, revealed quantitative
modifications of the results \cite{Manzke:02}. We show that for QD systems the situation is different  due to
the appearance of localized states with a discrete spectrum. The calculation of scattering integrals 
in terms of free-carrier energies breaks down for processes due to Coulomb interaction which involve only 
localized states. It turns out that these are among the dominant processes for QDs containing more
than one confined shell. For other processes involving localized states, their energy renormalization
is also of enhanced importance.
In the first part of this paper we present a non-Markovian treatment of dephasing 
due to Coulomb interaction 
where energy renormalizations under high-density conditions typical for QD lasers
are self-consistently included and the importance of various scattering channels is analyzed.

For the interaction of QD carriers with LA-phonons, extensive
work has been devoted to the low temperature regime,
\cite{Uskov:00,Krummheuer:02,Muljarov:04}. It is generally
acknowledged that acoustic phonons dominate the dephasing at low
temperatures. Nevertheless, several low temperature PL measurements
\cite{Htoon:01,Oulton:03} revealed LO-phonon replicas. In
Ref.~\cite{Borri:01} it was suggested that for temperatures above 100K
the dephasing due to carrier-LO-phonon interaction is of growing
importance. 
The interaction of QD carriers with LO phonons is strongly influenced by hybridization effects \cite{Inoshita:97},
which require the application of the polaron picture. A quantum-kinetic description of carrier-phonon
scattering based on polarons has recently been used to explain 
ultrafast carrier capture and relaxation processes in QDs \cite{Seebeck:05}.
In the second part of this paper, we present the corresponding treatment of dephasing processes. 
Then the combined influence of Coulomb interaction and carrier-phonon interaction
on optical absorption and gain spectra is analyzed for quasi-equilibrium excitation conditions.

In Refs.~\cite{Bayer:02} and \cite{Matsuda:03} PL spectroscopy
measurements revealed homogeneous linewidths of several meV at room
temperature. Comparable results have also been found by four-wave
mixing experiments \cite{Borri:01}.  Our microscopic calculation can
reproduce these experimental findings even though a direct
quantitative comparison, which is not the purpose of this paper, would
require better knowledge of the specific QD parameters as well as
possible improvements of our QD model.

The primary focus of this work is on QD optical spectra in the
room-temperature high-density regime relevant for laser applications.
For this situation, a quantum kinetic theory for carrier-carrier
interaction and carrier scattering with LO-phonons is developed. It is
shown that non-Markovian effects and energy renormalizations play an
essential role in QD systems.  The theory is evaluated to compare the
contributions of various scattering channels to dephasing and
line-shifts as well as their excitation density dependence. Studies of
the temperature dependence are restricted to T $\ge$ 150K where the
inclusion of LA-phonons is not necessary. 

Our investigations show that the efficiency of 
dephasing processes depends strongly on the carrier density in the system. 
For low carrier densities and room temperature, the electron-LO-phonon interaction is the dominant mechanism, leading to 
the appearance of additional side-peaks in the optical spectra due to polaronic hybridization effects.
Even though
for high carrier densities the Coulomb interaction becomes 
the dominant mechanism, the electron-LO-phonon interaction 
remains important as it continues to contribute to the linewidth and lineshape of the QD resonances.

For typical InGaAs QD parameters, a QD density of 
$10^{10}\text{cm}^{-2}$ on the WL and carrier densities around $10^{10}\text{cm}^{-2}$ at room temperature,
we obtain a homogeneous 
linewidth of the ground state resonance with a full width at half maximum (FWHM) of 3.2meV in 
agreement with the experimental findings of Ref.~\cite{Bayer:02}.
For lower carrier densities and decreasing temperature,
we can reproduce the observed dephasing reduction while
for elevated carrier densities and decreasing temperature,
the reduction of the dephasing due to LO-phonons is compensated 
by increasing dephasing due to Coulomb scattering.

\section{Theory of optical absorption \& gain calculations}\label{sec:theory}
The coherent optical excitation of semiconductors can be described in terms of interband transition amplitudes
$\psi_\alpha=\langle e_\alpha h_\alpha \rangle$ and population functions
$f^e_\alpha=\langle e^\dag_\alpha e_\alpha\rangle$
and $f^h_\alpha=\langle h^\dag_\alpha h_\alpha \rangle$ for electrons and holes, respectively.
For the annihilation and creation operators of electrons and holes, $\alpha$ combines the quantum numbers for 
either WL or QD states. The dynamics of $\psi_\alpha(t)$ and $f^{e,h}_\alpha(t)$ 
in response to the optical field $E(t)$
is determined by the SBE \cite{Haug:94} which can be used to incorporate 
many-body interaction effects to describe dephasing, energy renormalization and screening as 
demonstrated for QWs in Ref.~\cite{Jahnke:97}.
The optical absorption or gain coefficient can be defined via the response of the system to a weak optical field 
$E(\omega)$ in terms of the linear optical susceptibility $\chi(\omega)=P(\omega)/E(\omega)$. 
Here $P(\omega$) is the Fourier transform of the macroscopic polarization $P(t)=\sum_\alpha d^\ast_\alpha
\psi_\alpha$ with the interband dipole matrix element $d_\alpha$. 

A solution linear in the optical probe field $E(t)$ implies that changes of the electron and hole population 
$f^{e,h}_\alpha$ due to $E(t)$ can be neglected. In the following we consider quasi-equilibrium 
distributions for electron and holes which are approximately realized in injection-current driven
devices. Alternatively the carriers can be generated with an optical pre-pulse where a sufficient
delay of the probe pulse ensures the the excited carriers are thermalized and that coherent
effects of the pump pulse can be neglected.

In this regime, the Fourier transform of the coherent interband transition amplitude $\psi_\alpha$ 
obeys the equation

\begin{equation}\label{eq:sbe}
\begin{split}
 (\hbar\omega-\varepsilon^{e,\text{HF}}_\alpha-\varepsilon^{h,\text{HF}}_\alpha)&\psi_{\alpha}(\hbar\omega)
+\left[1-f^e_{\alpha}- f^h_{\alpha} \right ]\Omega^\text{HF}_\alpha(\hbar\omega)\\
 =&\ S^\text{Coul}_\alpha(\hbar\omega)+S^\text{Phon}_\alpha(\hbar\omega)~.
 \end{split}
\end{equation}

The single particle energies entering Eq.~\eqref{eq:sbe} include already renormalizations
due to Hartree-Fock (HF) Coulomb effects
according to $\varepsilon^{a,\text{HF}}_\alpha=e^a_\alpha+\Sigma^{a,\text{HF}}_\alpha$
with $a=e,h$. Here $e^a_\alpha$ are the free-carrier energies and 
{$\Sigma^{a,\text{HF}}_\alpha=
\sum_\beta\left(V_{\alpha\beta\beta\alpha}-V_{\alpha\beta\alpha\beta}\right)f^a_\beta$}
combines the Hartree and exchange Coulomb self-energy.
In a spatially homogeneous system like bulk semiconductors or for the in-plane motion for 
QWs, the Hartree terms cancel due to local charge neutrality. In the QD-WL system, the absence of local
charge neutrality leads to Hartree contributions for the QD states.
The influence of the WL states can be incorporated approximately within screened QD Hartree contributions
as discussed in detail in Ref.~\cite{Nielsen:05}. The Coulomb exchange contributions
to the interband Rabi energy {$\Omega^\text{HF}_\alpha(\hbar\omega)=
\vec{d}\cdot\vec{E}(\hbar\omega)+\sum_\beta V_{\alpha\beta\alpha\beta}\psi_\beta(\hbar\omega)$}
give rise to the excitonic resonance of the WL and QD states.

Without the terms $S^\text{Coul}_\alpha(\hbar\omega)$ and $S^\text{Phon}_\alpha(\hbar\omega)$
we would recover the well-known semiconductor Bloch equations \cite{Haug:94} 
formulated in an arbitrary eigenfunction basis.
$S^\text{Coul}_\alpha(\hbar\omega)$ and $S^\text{Phon}_\alpha(\hbar\omega)$ 
are used to describe correlation contributions which lead to 
dephasing of the coherent polarization and to the corresponding additional energy renormalizations due 
to Coulomb interaction and carrier-phonon interaction, respectively. 
These terms will be discussed in the following sections.
A general feature of the following description 
is that the dephasing contributions for both the Coulomb interaction and the carrier-phonon
interaction can be separated into a part which is diagonal in the state index $\alpha$ 
for the transition amplitude (diagonal dephasing $\Gamma^\text{DD}$)
and a corresponding off-diagonal part ($\Gamma^\text{OD}$) which mixes the coherent 
interband transition amplitudes for various states according to
\begin{equation}\label{eq:diagoff}
\begin{split}
S_\alpha(\hbar\omega)=&
-\Gamma^\text{DD}_\alpha(\hbar\omega)\psi_\alpha(\hbar\omega)\\
&+\sum\limits_{\alpha_1}\Gamma^\text{OD}_{\alpha\alpha_1}(\hbar\omega)\psi_{\alpha_1}(\hbar\omega)~.
\end{split}
\end{equation}
The imaginary part of $\Gamma^\text{DD}$ describes a dephasing (damping) of the coherent
polarization  which can be expressed as a state and frequency dependend $T_2$ time.
The contribution of $\Gamma^\text{DD}$ is however partly compensated by 
the off-diagonal contribution $\Gamma^\text{OD}$. The real parts of $\Gamma^\text{DD}$ and 
$\Gamma^\text{OD}$ give rise to additional renormalizations of the 
energies on the L.H.S. of Eq.~\eqref{eq:sbe}.

\section{Coulomb interaction}\label{sec:Coul}

\subsection{Theory}\label{sec:coultheory}
The correlation contributions due to Coulomb interaction, $S^\text{Coul}_\alpha(\hbar\omega)$,
are evaluated in the second-order Born approximation. To close the set of equation for the interband transition 
amplitude $\psi$ we additionally apply the generalized Kadanoff-Baym ansatz 
(GKBA) \cite{Lipa:86,Tso:91}.
Within this approach the correlation contributions due to Coulomb interaction read 
\begin{widetext}
\begin{eqnarray}\label{eq:sddcoul}
\Gamma^\text{DD}_\alpha(\hbar\omega)&=&i \sum\limits_{\substack{
a,b=e,h\\
b\neq a}}\,
\sum\limits_{\alpha_1\alpha_2\alpha_3}\\
& \times \Big \{& \nonumber 
W_{\alpha\alpha_2\alpha_3\alpha_1}
\left[ 2W^\ast_{\alpha\alpha_2\alpha_3\alpha_1}
- W^\ast_{\alpha\alpha_2\alpha_1\alpha_3} \right ]
g\left(\hbar\omega-\tilde\varepsilon^{b}_\alpha-\tilde\varepsilon^a_{\alpha_1}+
\left(\tilde\varepsilon^a_{\alpha_2}\right)^\ast-\tilde\varepsilon^a_{\alpha_3}\right)
\Big[\left(1-f^a_{\alpha_2}\right)f^a_{\alpha_3}f^a_{\alpha_1}+(f\rightarrow 1-f)\Big]\\[0.1cm]
&+ & \nonumber
2W^\ast_{\alpha\alpha_2\alpha_3\alpha_1}W_{\alpha\alpha_2\alpha_3\alpha_1}
\qquad\qquad\qquad\;\; g\left(\hbar\omega-\tilde\varepsilon^{b}_\alpha-\tilde\varepsilon^a_{\alpha_1}-
\tilde\varepsilon^{b}_{\alpha_2}+\left(\tilde\varepsilon^{b}_{\alpha_3}\right)^\ast\right)
\Big[f^{b}_{\alpha_2}\left(1-f^{b}_{\alpha_3}\right)f^a_{\alpha_1}+(f\rightarrow 1-f)\Big]\Big\}~,
\end{eqnarray}
and 
\begin{eqnarray}\label{eq:sodcoul}
\Gamma^\text{OD}_{\alpha\alpha_1}(\hbar\omega)&=&i \sum\limits_{\substack{
a,b=e,h\\
b\neq a}}\,
\sum\limits_{\alpha_2\alpha_3}\\
&\times \Big \{ & \nonumber 
W_{\alpha\alpha_2\alpha_3\alpha_1} 
\left[ 2W^\ast_{\alpha\alpha_2\alpha_3\alpha_1}
-W^\ast_{\alpha\alpha_2\alpha_1\alpha_3} \right ]
g\left(\hbar\omega-\tilde\varepsilon^a_\alpha-\tilde\varepsilon^{b}_{\alpha_1}
-\tilde\varepsilon^a_{\alpha_2}+\left(\tilde\varepsilon^a_{\alpha_3}\right)^\ast\right)
\Big[\left(1-f^a_{\alpha_3}\right)f^a_{\alpha_2}f^a_{\alpha}+(f\rightarrow 1-f)\Big]\\
&+ & \nonumber
2W_{\alpha\alpha_2\alpha_3\alpha_1}  W^\ast_{\alpha\alpha_2\alpha_3\alpha_1}
\qquad\qquad\qquad\;\; g\left(\hbar\omega-\tilde\varepsilon^a_\alpha-\tilde\varepsilon^{b}_{\alpha_1}
+\left(\tilde\varepsilon^{b}_{\alpha_2}\right)^\ast-\tilde\varepsilon^{b}_{\alpha_3}\right)
\Big[f^{b}_{\alpha_3}\left(1-f^{b}_{\alpha_2}\right)f^a_{\alpha}+(f\rightarrow 1-f)\Big]\Big\}~,
\end{eqnarray}
\end{widetext}
with the screened Coulomb interaction matrix elements $W_{\alpha\alpha_2\alpha_3\alpha_1}$ 
discussed in Appendix \ref{sec:qdmodel}.
Note that $\tilde\varepsilon^a_\alpha=\varepsilon^a_\alpha-i\gamma^a_\alpha$ 
appearing in the function $g(\Delta)=\frac{i}{\Delta}$
are effective complex single-particle energies which combine renormalized energies $\varepsilon^a_\alpha$ as 
well as the corresponding quasi-particle damping $\gamma^a_\alpha$. 
This is elaborated in more detail below.

As for the particle scattering \cite{Nielsen:04} we have direct
and exchange contributions which are proportional to $2|W_{\alpha\alpha_2\alpha_3\alpha_1}|^2$
and $W_{\alpha\alpha_2\alpha_3\alpha_1}W^\ast_{\alpha\alpha_2\alpha_1\alpha_3}$, respectively.
The population factors describe the availability of initial and final states. For the above discussed excitation
conditions, the population factors are time independent and 
we have restricted ourselves to the contributions linear in the transition amplitude $\psi_\alpha$.

The frequency dependence of $\Gamma^\text{DD}$ and $\Gamma^\text{OD}$ in Eqs.~\eqref{eq:sddcoul} 
and \eqref{eq:sodcoul} reflect the non-Markovian treatment of the dephasing processes. Indeed, in this
case, the product of the type $\Gamma\psi$ of Eq.~\eqref{eq:diagoff} amounts, in the time domain, to a convolution
integral describing memory effects. The Markovian limit is obtained by pulling out from this integral the slowly
varying component of $\psi_\alpha(t)$. Then the fast component 
$e^{\frac{i}{\hbar}(
\tilde\varepsilon^a_{\alpha}+
\tilde\varepsilon^b_{\alpha})t}$ ($a\neq b$) fixes the frequency value at 
$\omega=\omega_{\alpha}=\frac{1}{\hbar}(\tilde\varepsilon^e_{\alpha}+
\tilde\varepsilon^h_{\alpha})$ for $\Gamma^\text{DD}$ and $\omega=\omega_{\alpha_1}$ for $\Gamma^\text{OD}$,
so that in this limit both dephasing contributions become 
frequency independent (see Appendix \ref{sec:memory} for details).
If one combines the Markov approximation with the use of free-carrier energies
in the scattering contributions, the diagonal dephasing is given 
by the sum of in- and out-scattering rates, 
$\text{Im}(\Gamma^\text{DD}_\alpha)=S^\text{in}_\alpha+S^\text{out}_\alpha$.
Here $S^\text{in,out}_\alpha$ are the rates in the Markovian kinetic equation 
for the carrier population, as defined 
e.g. in Ref.~\cite{Nielsen:04}. 
Results of these equations for optical spectra have been studied in detail for QW systems. If one restricts
the analysis to diagonal dephasing contributions $\Gamma^\text{DD}$ and neglects off-diagonal dephasing terms
$\Gamma^\text{OD}$, damping of the excitonic resonances is grossly overestimated \cite{Jahnke:97}. 
Non-Markovian calculations further reduce the interaction-induced broadening and line-shift of the excitonic
resonances in QW systems \cite{Manzke:02}.

For the following discussion we would like to point out that one has to compare on one side the Markov 
approximation with a non-Markovian treatment and on the other side the use of free-carrier energies versus 
renormalized energies in the scattering integrals.
The non-Markovian treatment is more crucial in QD systems compared to QWs, since the discrete part of the spectrum 
with large energy separation emphasizes the frequency-dependence of the dephasing rates \cite{Schneider:04}.

For the use of free-carrier energies in the scattering contributions, the limit 
$\gamma^a_\alpha\rightarrow 0, \gamma^a_\alpha>0$ leads to 
$g(\Delta)=\pi\delta(\Delta)+i P \frac{1}{\Delta}$ where $P$ denotes 
the principal value integral. This approximation implies
serious difficulties in a QD system when processes are taken into account which involve only
discrete states. In these cases the $\delta$-functions are not integrated out and 
thus the results are not well defined. 
Note that this is not a artifact of the Markov approximation but also applies to the non-Markovian
calculation.
Even if one would introduce a finite broadening of the $\delta$-function, e.g., due to interaction of carriers
with acoustic phonons, the non-Markovian calculation still yields unphysical results. 
This is because the spectrum predicted by Eq.~\eqref{eq:sbe} has peaks at the energies of the L.H.S.
renormalized by the correlation contributions on the R.H.S. and 
%
%
hence, 
$\hbar\omega$ samples renormalized interband transitions. If they are mixed with free-carrier energies
in the correlation contributions Eqs.~\eqref{eq:sddcoul} and \eqref{eq:sodcoul}, via $g(\hbar\omega-\Delta)$,
then
artificially also free carrier transitions and hybridization effects between all these energies 
can appear in the optical spectra,
which are absent when self-consistently renormalized energies are used.

Another important point, which complicates the treatment of the coupled QD-WL system, is the existence 
of many scattering channels which are often partly neglected (for a discussion of the scattering channels 
in the framework of carrier kinetics see Ref.~\cite{Nielsen:04}).
This can strongly influence the results as we will show in Section \ref{sec:resCoul}.


\subsubsection*{Self-consistent single-particle energy renormalizations}
The renormalization of the single-particle energies, which enter in the dephasing rates, originates
from the same interaction processes that determine the dephasing itself. Technically, they can be traced back to 
the same many-body self-energy. The renormalized energies can be obtained from the poles of the retarded
Green's function (GF) in Fourier space. In the absence of interaction, the retarded GF is given by 
\begin{equation}\label{eq:freeGret}
G^\text{0,ret}_{a,\alpha}(\hbar\omega)=\frac{1}{\hbar\omega-e^a_\alpha+i\delta}
\end{equation}
with the single-particle energy $e^a_\alpha$ and $\delta\rightarrow 0,\delta>0$. At elevated carries densities,
it is a reasonable approximation, that the main effect of the Coulomb interaction is a shift of the 
single-particle energy and the addition of a quasi-particle broadening such that the single-pole
structure of the retarded GF remains valid. This picture corresponds to the Landau theory of a Fermi liquid
and leads to the retarded GF of the interacting system
\begin{equation}
\begin{split}\label{eq:renGret}
G^\text{ret}_{a,\alpha}(\hbar\omega)&=\frac{1}{\hbar\omega-\tilde\varepsilon^a_\alpha}
=\frac{1}{\hbar\omega-\varepsilon^a_\alpha+i\gamma^a_\alpha }~.
\end{split}
\end{equation}
With this ansatz, we define a complex effective single-particle energy
$\tilde\varepsilon^a_\alpha=\varepsilon^a_\alpha-i\gamma^a_\alpha$ which 
consists of a renormalized energy $\varepsilon^a_\alpha$ and the corresponding 
quasi-particle damping $\gamma^a_\alpha$.
In the pole approximation, the self-consistency requirement leads to
\begin{equation}\label{eq:selfcons}
\tilde\varepsilon^a_\alpha=
e^a_\alpha+
\Sigma^{a,\text{HF}}_\alpha+
\Sigma^{a,\text{ret}}_\alpha(\tilde\varepsilon^a_\alpha)
\end{equation}
with the retarded self-energy
\begin{widetext}
\begin{eqnarray}\label{eq:sigma_ret}
\Sigma^{a,\text{ret}}_\alpha(\hbar\omega)&=&-i~\sum\limits_{\substack{
b=e,h\\
b\neq a}}\,
\sum\limits_{\alpha_1\alpha_2\alpha_3}\\
&\times \Big \{ & \nonumber
W^\ast_{\alpha\alpha_2\alpha_3\alpha_1}\left[ 2W^\ast_{\alpha\alpha_2\alpha_3\alpha_1}
- W^\ast_{\alpha\alpha_2\alpha_1\alpha_3} \right ]
g\left(\hbar\omega-\tilde\varepsilon^a_{\alpha_1}+
\left(\tilde\varepsilon^a_{\alpha_2}\right)^\ast-\tilde\varepsilon^a_{\alpha_3}\right)
\Big[\left(1-f^a_{\alpha_2}\right)f^a_{\alpha_3}f^a_{\alpha_1}+(f\rightarrow 1-f)\Big]\\[0.1cm]
& & \nonumber
+2W^\ast_{\alpha\alpha_2\alpha_3\alpha_1}W_{\alpha\alpha_2\alpha_3\alpha_1}
\qquad\qquad\qquad
g\left(\hbar\omega-\tilde\varepsilon^a_{\alpha_1}-\tilde\varepsilon^{b}_{\alpha_2}+
\left(\tilde\varepsilon^{b}_{\alpha_3}\right)^\ast\right)
\Big[f^{b}_{\alpha_2}\left(1-f^{b}_{\alpha_3}\right)f^a_{\alpha_1}+(f\rightarrow 1-f)\Big]\Big\}~.
\end{eqnarray}
\end{widetext}
The close connection between the diagonal dephasing and the retarded 
self-energy $\Sigma^{a,\text{ret}}_\alpha$ can be expressed as
\begin{equation}
\Gamma^\text{DD}_\alpha(\hbar\omega)=
\sum\limits_{\substack{
a,b=e,h\\
b\neq a}}\,\Sigma^{a,\text{ret}}_\alpha(\hbar\omega-\tilde\varepsilon^{b}_\alpha)~.
\end{equation}

\subsection{Results}\label{sec:resCoul}
Throughout this paper, results are presented for a carrier and lattice 
temperature of 300K and a density of QDs on the WL of $n_\text{QD}=10^{10}\,\text{cm}^{-2}$. 
Further details of the QD model are given in Appendix \ref{sec:qdmodel}.

In Fig.~\ref{fig:coul_dens} we show optical absorption spectra for the combined QD-WL system 
and various carrier densities. 
Excitation-induced dephasing and energy renormalizations due to Coulomb interaction have been
included according to Eqs.~\eqref{eq:sddcoul},\eqref{eq:sodcoul} and \eqref{eq:selfcons},\eqref{eq:sigma_ret}.
Identifiable are the excitonic resonance of 
the WL at around \mbox{-15meV}
as well as the p-shell and s-shell resonances at about \mbox{-90meV} and \mbox{-150meV},
respectively.
\begin{figure}[!ht]
    \begin{center}
    \includegraphics*[width=.45\textwidth]{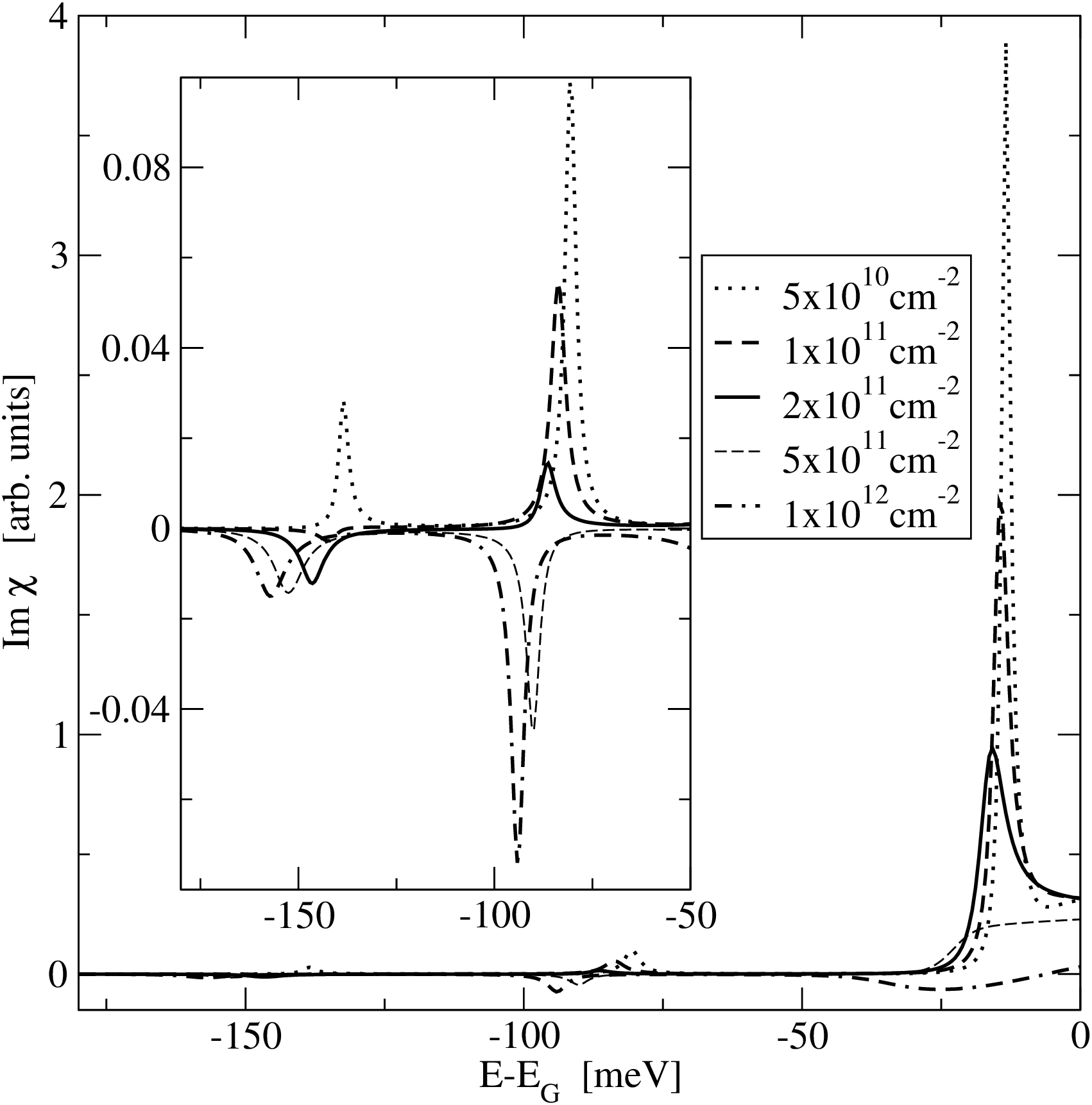}	
    \caption{Imaginary part of the optical susceptibility for the combined QD-WL 
    		system including interaction-induced dephasing and line shifts due to Coulomb interaction 
		for various total carrier densities. The inset shows a 
	 	scale up of the QD resonances.
	 	\label{fig:coul_dens}
	 }
    \end{center}
\end{figure}
We observe a strong damping of these resonances with increasing carrier density 
which is accompanied by a pronounced red shift of the QD lines. The transition from absorption to gain
takes place at a density $\sim \dens{1}{11}$ ($\sim \dens{5}{11}$) for the s-shell (p-shell).
An important point for practical applications is the saturation of the optical gain for the s-shell
accompanied by an increasing red-shift.
This is due to a combination of state filling and saturation of dephasing at the s-shell resonance.
In contrast the gain at the p-shell resonance increases further and shows no saturation 
for the densities investigated here.


\subsubsection*{Importance of different dephasing processes and frequently used approximations}
The Coulomb matrix elements allow to distinguish between different dephasing processes
in complete analogy to the carrier scattering \cite{Nielsen:04}. 
If all four indices of $W_{\alpha\alpha_2\alpha_3\alpha_1}$ 
are WL states, we refer to the corresponding processes as WL relaxation. They describe dephasing and energy shifts 
of the WL states in the optical spectra.
Apart from the fact that we have to include OPW corrections to the interaction
matrix elements for a proper description of the coupled QD-WL system \cite{Nielsen:04}, 
this resembles the case of a QW system.
Since the processes involving QD states (except intra-dot processes, see below) are discussed
in great detail in Ref.~\cite{Nielsen:04} we will only give a brief survey here.

If one of the four indices of $W_{\alpha\alpha_2\alpha_3\alpha_1}$ corresponds to a QD state, we 
consider the dephasing mechanism as
WL assisted carrier-capture, because in the corresponding scattering process
an electron or a hole is captured from the WL to the 
QD \cite{Nielsen:04}. 
Likewise we refer to processes with two QD state indices as WL assisted relaxation and
to processes with three QD state indices as dot-assisted processes.
Intra-dot processes are described when all four indices of $W_{\alpha\alpha_2\alpha_3\alpha_1}$ correspond to QD 
states. These scattering events are not important in the carrier dynamics, 
because we only consider two confined shells for electrons and holes and therefore such 
scattering processes cannot redistribute carriers. 
However, they clearly provide additional dephasing of the coherent polarization.
Note that the intra-dot processes are not the only 
events which produce dephasing without redistribution of carriers.
Such processes, in which two carriers switch their states, also appear in the WL assisted relaxation 
and in the WL relaxation.

\begin{figure}[!ht]
    \begin{center}
	\includegraphics*[width=.45\textwidth]{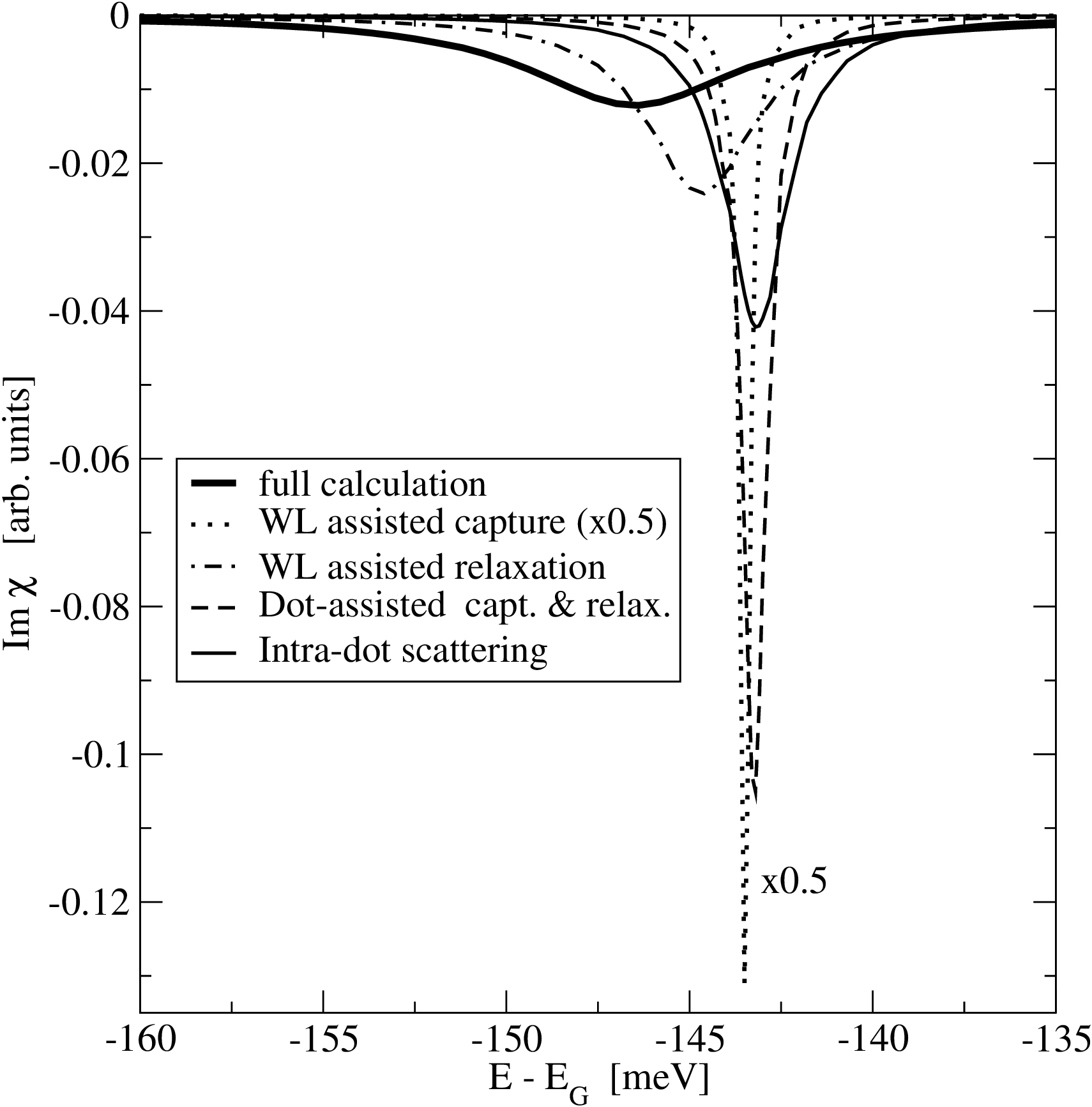}	
	\caption{Imaginary part of the optical susceptibility for the s-shell resonances
		at a carrier density of $\dens{2}{11}$. The result including all considered Coulomb scattering 
		processes (thick solid line) is compared to calculations where only certain classes of processes 
		are evaluated.
		\label{fig:coul_processes}
		}
    \end{center}
\end{figure}
Figure \ref{fig:coul_processes} shows that WL assisted relaxation (dash-dotted line) and intra-dot scattering
(thin solid line), which include the processes leading to dephasing without redistribution of carriers,
are the most important contributions, while WL assisted capture (dotted line) and dot-assisted (dashed line)
processes cause a rather small dephasing. However, since this picture can change with carrier density,
it is important to test the influence of all dephasing channels.

\begin{figure}[!ht]
    \begin{center}
    \includegraphics*[width=.45\textwidth]{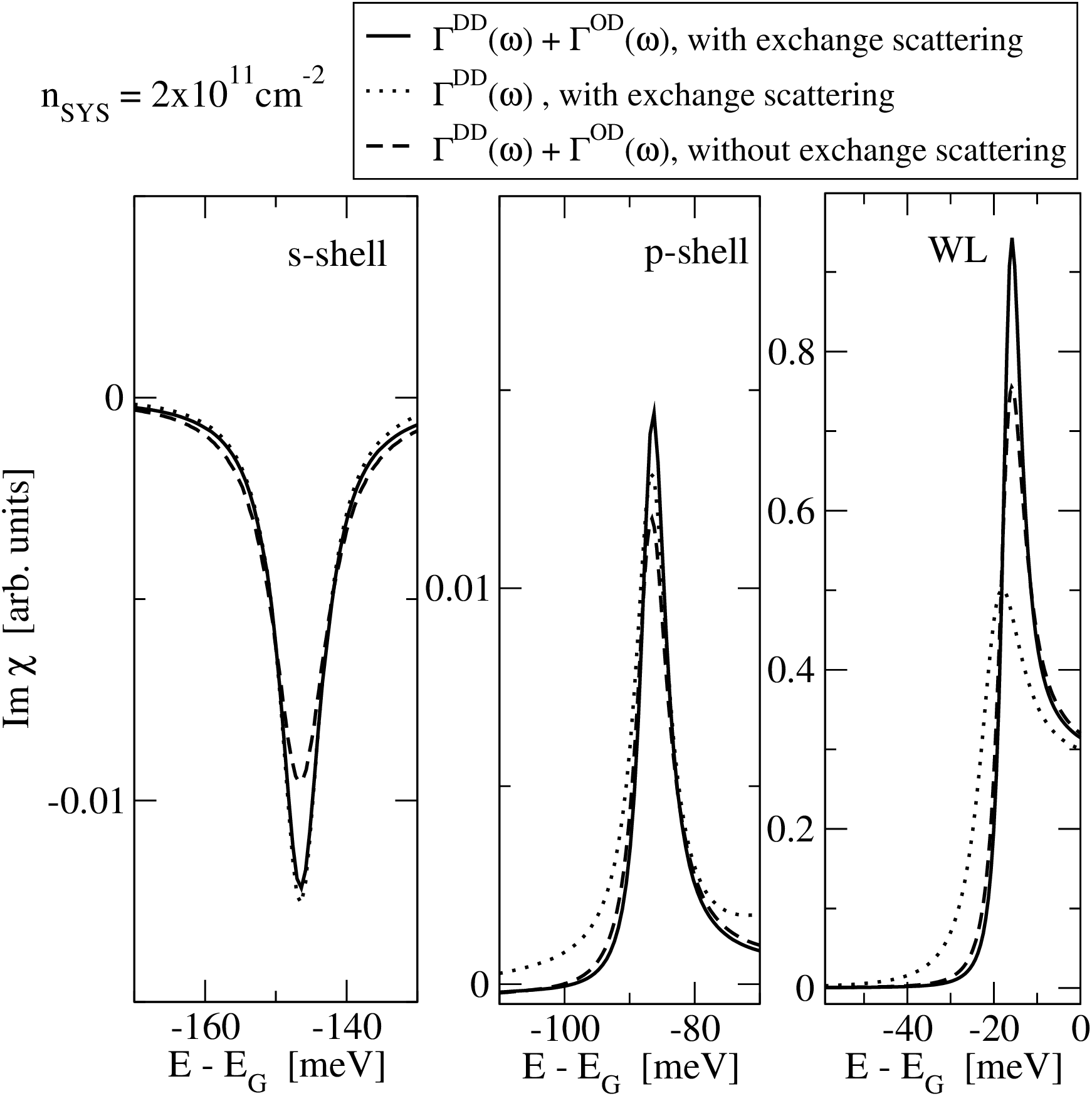}
    \caption{Imaginary part of the optical susceptibility for a carrier density of $\dens{2}{11}$
    	with full dephasing contributions (solid lines), with diagonal dephasing contributions (dotted lines),
	and with diagonal and off-diagonal dephasing but without the exchange scattering in both terms 
	(dashed lines).
	 \label{fig:coul_DD}
	 }
    \end{center}
\end{figure}

In Fig. \ref{fig:coul_DD} we compare the optical susceptibility from the full calculation (solid line) 
with a result where
only the diagonal dephasing contributions were taken into account (dotted lines).
For the excitonic resonance of the WL,
the absence of the off-diagonal dephasing leads to a overestimation of the linewidth by 
roughly a factor of two under the considered high-density conditions.
Using only the diagonal dephasing contributions turns out to be a reasonable approximation for 
the lowest QD resonances while for the excited QD transition the lineshape is not fully reproduced. Off-diagonal
dephasing contributions are less important for the QD resonances due to
the rather large spectral separation between the QD states and the WL states and for the same reason
off-diagonal contributions are stronger for the p-shell than for the s-shell.

Note that the foregoing discussion applies to the non-Markovian treatment.
If one neglects the off-diagonal dephasing and
uses the Markov approximation, again the dephasing is grossly overestimated when all relevant scattering processes
are included. For the discussed QD model, transitions due to localized states are completely
damped out in this approximation.

The influence of the exchange term in the dephasing rates, which is often disregarded,
is also investigated in Fig.~\ref{fig:coul_DD}.
Neglecting the exchange scattering in the dephasing contributions clearly overestimates
the homogeneous linewidth by about 30 \%, thus pointing out that the exchange terms are almost as important
as the direct term and should be included in the calculation.


\subsubsection*{Renormalization of single-particle states}
\begin{figure}[!ht]
\vspace*{0.5cm}
    \begin{center}
    \includegraphics[width=.48\textwidth]{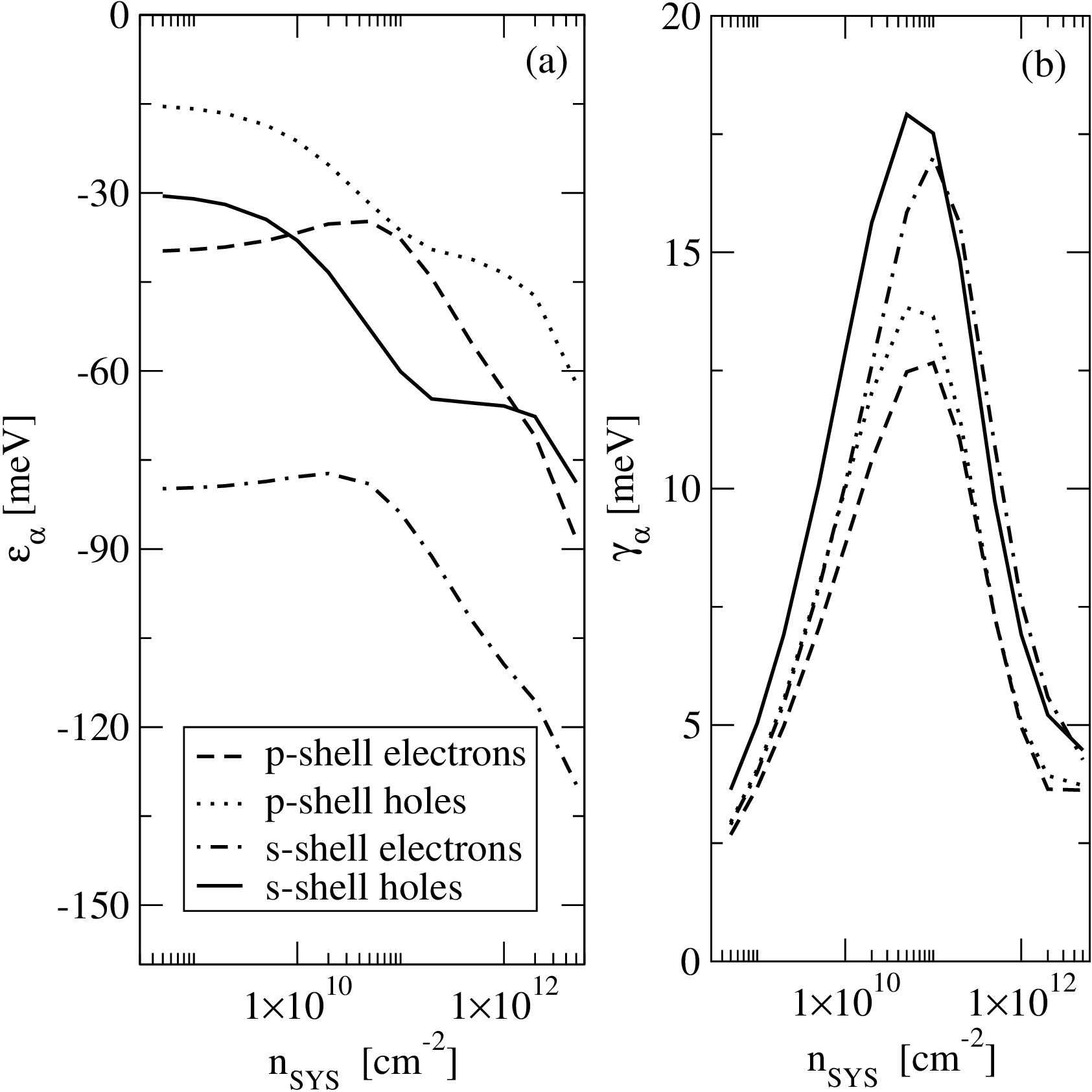}	
    \caption{
    Renormalized single-particle energies of the QD states for electrons and holes
    as a function of the total carrier density in the system (a) and corresponding single-particle broadening(b).
	 \label{fig:coul_energy}
	 }
    \end{center}
\end{figure}

As  discussed above, the single particle renormalizations
are of critical importance for the proper determination of the dephasing rates.
In Fig. \ref{fig:coul_energy}a the single particle renormalizations due to Coulomb interaction
are shown.
The Fock and correlation contributions lead to a decrease of the single-particle energies
with increasing carrier density. The Hartree terms increase the energies for electrons and decrease 
the energies for holes (for a detailed discussion cf. \cite{Nielsen:05}). This leads to 
slightly increasing (decreasing) single-particle energies 
for the electrons (holes) from low to intermediate carrier densities. 
For higher densities, both electron and hole energies strongly decrease.

With increasing carrier densities the single-particle broadening (Fig.~\ref{fig:coul_energy}b) 
strongly grows up to intermediate carrier densities
due to the larger availability of scattering partners. 
For higher carrier density, the strong single-particle energy shifts 
lead to a reduction of the scattering efficiency. 
Additionally at very high carrier densities the intra-dot processes
are suppressed due to Pauli-blocking. These effects result in a strong decrease of the single-particle
broadening.

\section{Carrier-phonon interaction}


\subsection{Theory}\label{sec:phontheory}
In this section, we compute correlation contributions in the optical spectra 
due to interaction of carriers with LO-phonons.
In contrast to the frequently applied time-dependent perturbation theory, it has been pointed out in 
Ref.~\cite{Inoshita:97} that the interaction involving localized QD 
carriers requires a description in the polaron picture.
In Ref.~\cite{Seebeck:05}, quasi-particle renormalizations described by a polaronic retarded GF have been used 
with in a quantum-kinetic theory to evaluate scattering processes and populations changes.
We extent this treatment to the polarization dynamics to analyze the corresponding dephasing and interband-energy 
renormalizations.

Within the random-phase approximation (RPA)
and the GKBA, the correlation contributions due to LO-phonons are
\begin{widetext}
\begin{eqnarray}\label{eq:sddphon}
\Gamma^\text{DD}_\alpha(\hbar\omega)&=&i\; \sum\limits_{\substack{
a,b=e,h\\
b\neq a}}\,
\sum\limits_{\beta}\frac{M^2_{LO}}{e^2/\varepsilon_0}V_{\alpha\beta\alpha\beta} \Big\{ 
\left(1-f^a_{\beta}\right)\big[
\left(1+N_\text{LO}\right)	G^{a,b}_{\beta,\alpha}(\hbar\omega-\hbar\omega_\text{LO})
+N_\text{LO}
\;\;G^{a,b}_{\beta,\alpha}(\hbar\omega+\hbar\omega_\text{LO})\big]\\
& & \nonumber 
\qquad\qquad\qquad\qquad\qquad\qquad
+\ f^a_{\beta}\ \big[
\left(1+N_\text{LO}\right)	G^{a,b}_{\beta,\alpha}(\hbar\omega+\hbar\omega_\text{LO})+
N_\text{LO}	\;\;
 G^{a,b}_{\beta,\alpha}(\hbar\omega-\hbar\omega_\text{LO})\big]\Big\} 
\end{eqnarray}
and 
\begin{eqnarray}\label{eq:sodphon}
\Gamma^\text{OD}_{\alpha\beta}(\hbar\omega)&=&i\; \sum\limits_{\substack{
a,b=e,h\\
b\neq a}}\,
\frac{M^2_{LO}}{e^2/\varepsilon_0}V_{\alpha\beta\alpha\beta} \Big\{ 
\left(1-f^a_{\alpha}\right)\big[
\left(1+N_\text{LO}\right)	G^{b, a}_{\beta,\alpha}(\hbar\omega-\hbar\omega_\text{LO})
+N_\text{LO}	\;\;	
G^{b, a}_{\beta,\alpha}(\hbar\omega+\hbar\omega_\text{LO})\big]\\
& & \nonumber 
\qquad\qquad\qquad\qquad\qquad\quad
+ \ f^a_{\alpha}\ \big[
\left(1+N_\text{LO}\right)	G^{b, a}_{\beta,\alpha}(\hbar\omega+\hbar\omega_\text{LO})+
N_\text{LO}	\;\;	
G^{b, a}_{\beta,\alpha}(\hbar\omega-\hbar\omega_\text{LO})\big]\Big\}
\end{eqnarray}
\end{widetext}
where the prefactor $M^2_\text{LO}=
4\pi\alpha \frac{\hbar}{\sqrt{2m}}
(\hbar\omega_\text{LO})^{3/2}$ includes the polar coupling strength
$\alpha$, the reduced mass $m$ and the phonon energy $\hbar\omega_\text{LO}$. 
In this equations we have introduced the abbreviation

\begin{equation}
G^\text{a,b}_{\beta,\alpha}(\omega)=\int d\tau \;e^{i\omega\tau}\;
G^{\text{a,ret}}_\beta(\tau)\; G^{\text{b,ret}}_\alpha(\tau)
\end{equation}
which represents the Fourier transform of a product of two retarded polaronic GFs.
Please note that the structure of this quantity can be 
traced back to the change from the $c,v$ picture to the $e,h$ picture. 
Through these functions polaronic renormalization effects such as phonon replicas and hybridization 
between the localized states \cite{Seebeck:05} are included in Eqs.~\eqref{eq:sddphon} 
and \eqref{eq:sodphon}. 
Details for the calculation of the retarded polaronic GF are give in Appendix \ref{sec:polGF}.


\subsection{Results}
\begin{figure}[!ht]
\vspace*{0.5cm}
    \begin{center}
    \includegraphics[width=.45\textwidth]{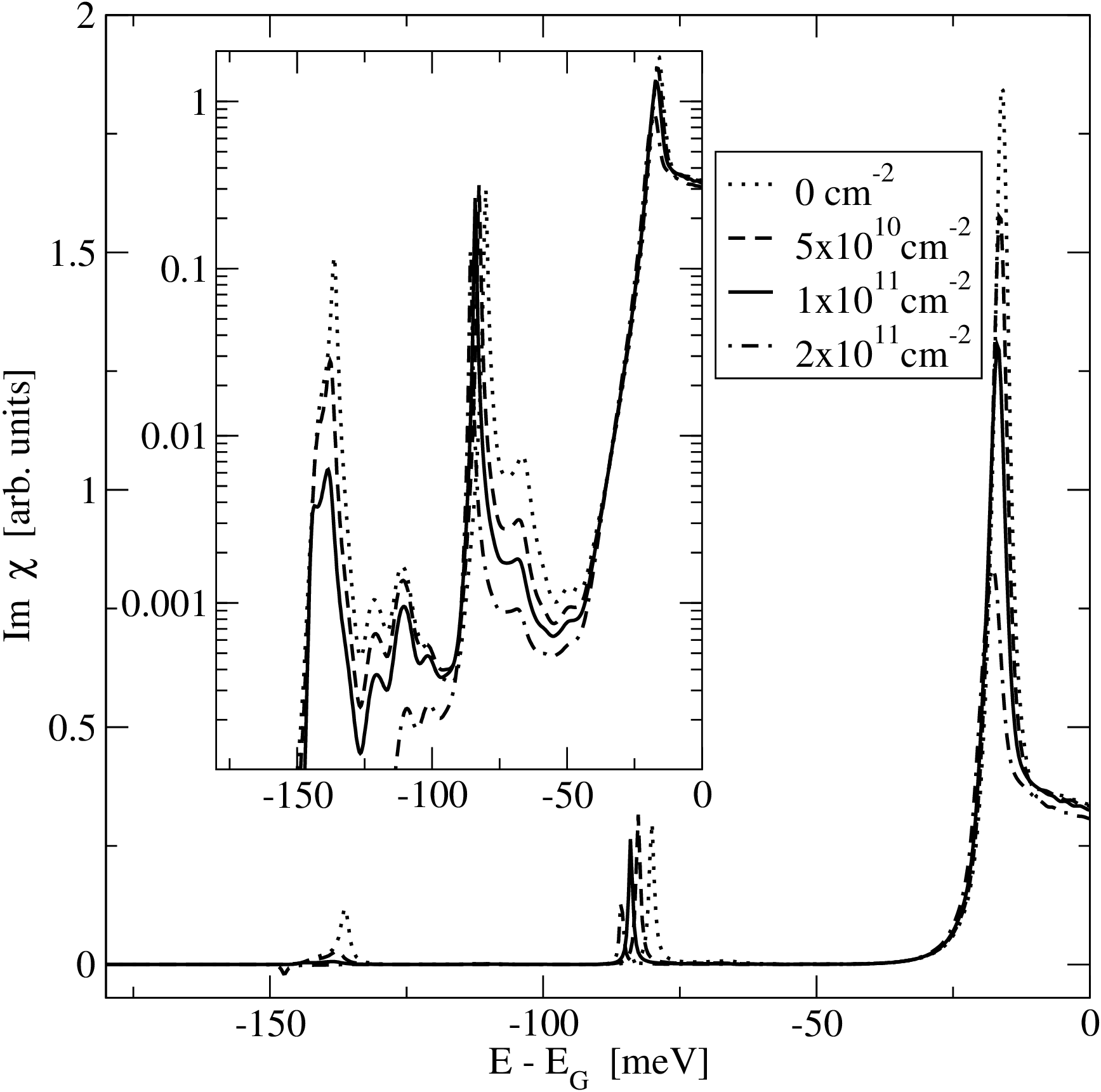}	
    \caption{
    		Imaginary part of the optical susceptibility for the combined QD-WL 
    		system including interaction-induced dephasing and line shifts due to carrier-phonon interaction 
		for various total carrier densities. 
    		The inset is a logarithmic plot of the QD resonances.
    	 \label{fig:phon_dens}
	 }
    \end{center}
\end{figure}

The high-density spectra with dephasing due to Coulomb 
interaction exhibit only three resonances, namely s-shell, p-shell 
and the excitonic resonance of the WL.
Absorption spectra with correlation contributions due to 
interaction of carriers with LO phonons are shown in Fig.~\ref{fig:phon_dens}. 
Polaronic renormalizations of the single-particle states lead to a more complicated resonance
structure for the interband transitions displayed in the inset of Fig.~\ref{fig:phon_dens}.
One can identify phonon replicas and results of the hybridization of the single-particle states. 
For example, the s-shell resonances 
has a shoulder on the lower energetic side 
due to hybridization of the corresponding electron state.
Energetically above the resonances of the s-shell and the p-shell several peaks 
due to phonon replicas and their hybridizations can be observed.
The non-Lorentzian character of the lineshapes
is even more pronounced as for the spectra with Coulomb interaction.
On the other hand, the damping of the resonances at higher densities remains weak. 


\section{Dephasing due to carrier-phonon and Coulomb interaction}\label{sec:coulphon}
In this section, we investigate the combined influence of the carrier-carrier Coulomb
scattering and the interaction of carrier with LO phonons. It turns out that polaronic
resonances in the single-particle spectral function are strongly damped out due to Coulomb scattering of carriers 
even at low carrier densities, so that a pole approximation for the single-particle properties
of the interacting system is reasonable. On the other hand, interband energy renormalizations
and dephasing contributions due to the interaction of carriers with LO phonons continue to be important.
Since our goal is a calculation of optical spectra for the QD-WL system under the influence of correlation effects 
at elevated carrier densities, 
we use for the description of single-particle properties a retarded GF obeying 
Eq.~\eqref{eq:Dyson_G_ret} with the free carrier energy 
in the L.H.S. replaced by $\tilde\varepsilon^a_\alpha=e^a_\alpha+\Delta^a_\alpha-i\gamma^a_\alpha$,
which is
determined by Eqs.~\eqref{eq:selfcons} and \eqref{eq:sigma_ret}. In other words the polaron is obtained by 
dressing with the phonon interaction not the free particles but the quasi-particles obtained by Coulomb 
renormalization.
Rewriting the retarded GF as 
\begin{equation}\label{eq:Gret}
G^{\text{a,ret}}_{\alpha}(\tau)=
\mathcal{G}^a_\alpha(\tau)\ 
e^{-\frac{i}{\hbar}(\Delta^a_\alpha-i\gamma^a_\alpha)\tau}~.
\end{equation}
we separate in $\mathcal{G}^a_\alpha(\tau)$ the phonon renormalization effects which are, of course,
influenced by the presence of the Coulomb interaction.
(The equation obeyed by $\mathcal{G}$ still contains $\Delta$ and $\gamma$.)
The finite life-time of these quasi-particles produces in general sufficient damping
to reduce the polaronic GF $\mathcal{G}^a_\alpha$ to a single-pole structure.
This pole is used instead of 
$e^a_\alpha$ in Eq.~\eqref{eq:selfcons}.
In this this way the Coulombian and the polaronic problems
become coupled and have to be solved self-consistently. 
The iterative solution to this problem converges rapidly.
\begin{figure}[!ht]
\vspace*{0.5cm}
    \begin{center}
    \includegraphics[width=.45\textwidth]{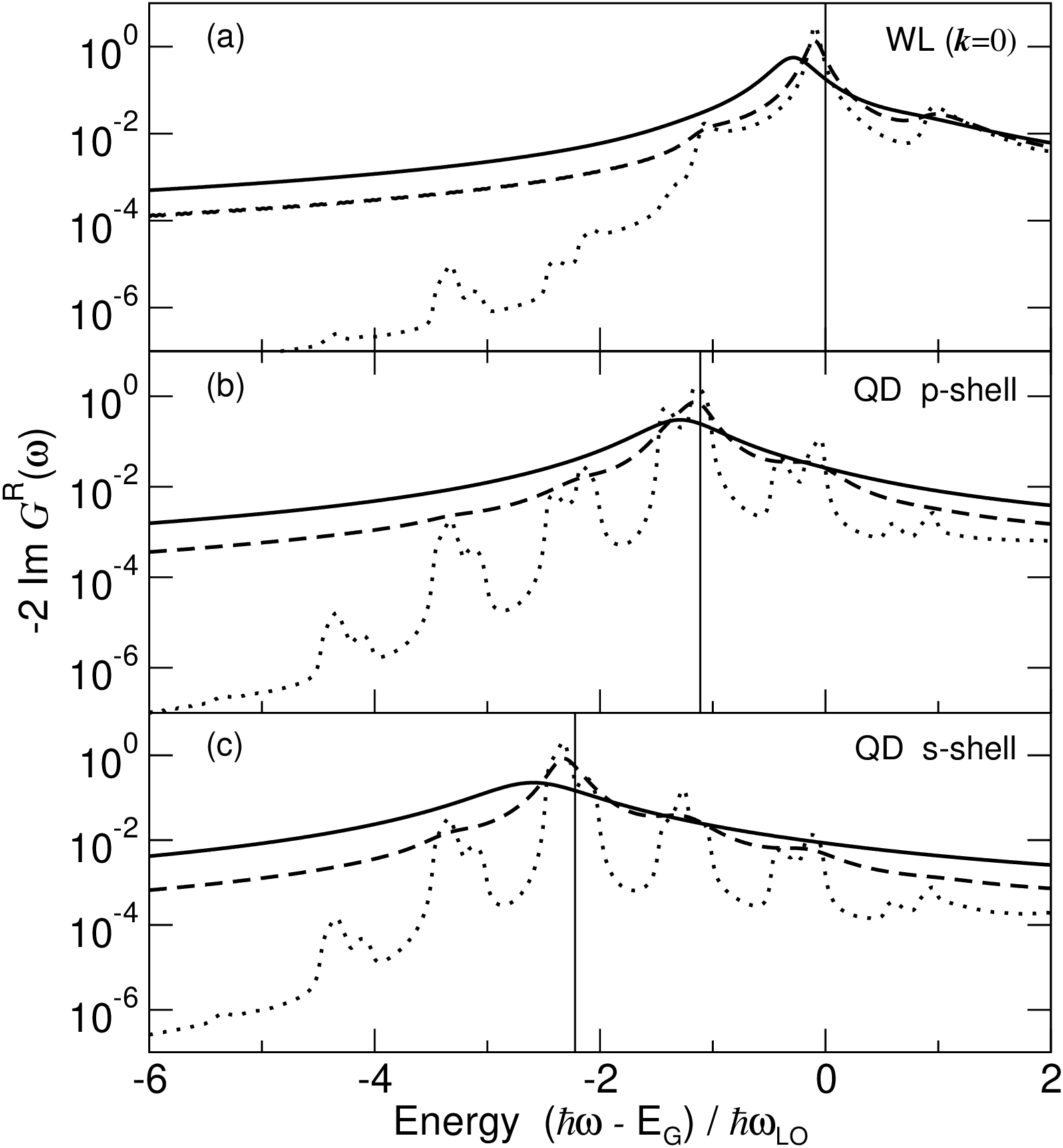}	
    \caption{Spectral function of the k=0 WL state and the QD p- and s-shell states including the interaction
		with LO phonons and Coulomb interaction of carriers. The total carrier density is 0 (dotted 
		line),	$\dens{5}{8}$ (dashed line) and $\dens{2}{11}$ (solid line). Vertical lines indicate
		the free carrier energies.
	 \label{fig:spec_func}
	 }
    \end{center}
\end{figure}
Results of the spectral function  of QD and WL states are shown in Fig.~\ref{fig:spec_func}. Even for low carrier 
densities, polaronic structures are strongly broadened as a result of the dominant role of damping due to Coulomb 
scattering. (For low carrier densities the exponential decay $e^{-\frac{\gamma^a_\alpha}{\hbar}\tau}$ due to 
Coulomb interaction, which is superimposed to the polaronic function $\mathcal{G}^a_\alpha(\tau)$ in 
Eq.~\eqref{eq:Gret}, might somewhat overestimate the damping of the polaronic resonances.
Nevertheless at higher carrier densities we expect a strong broadening of the polaron satellites.)

\begin{figure}[!ht]
\vspace*{0.5cm}
    \begin{center}
    \includegraphics[width=.45\textwidth]{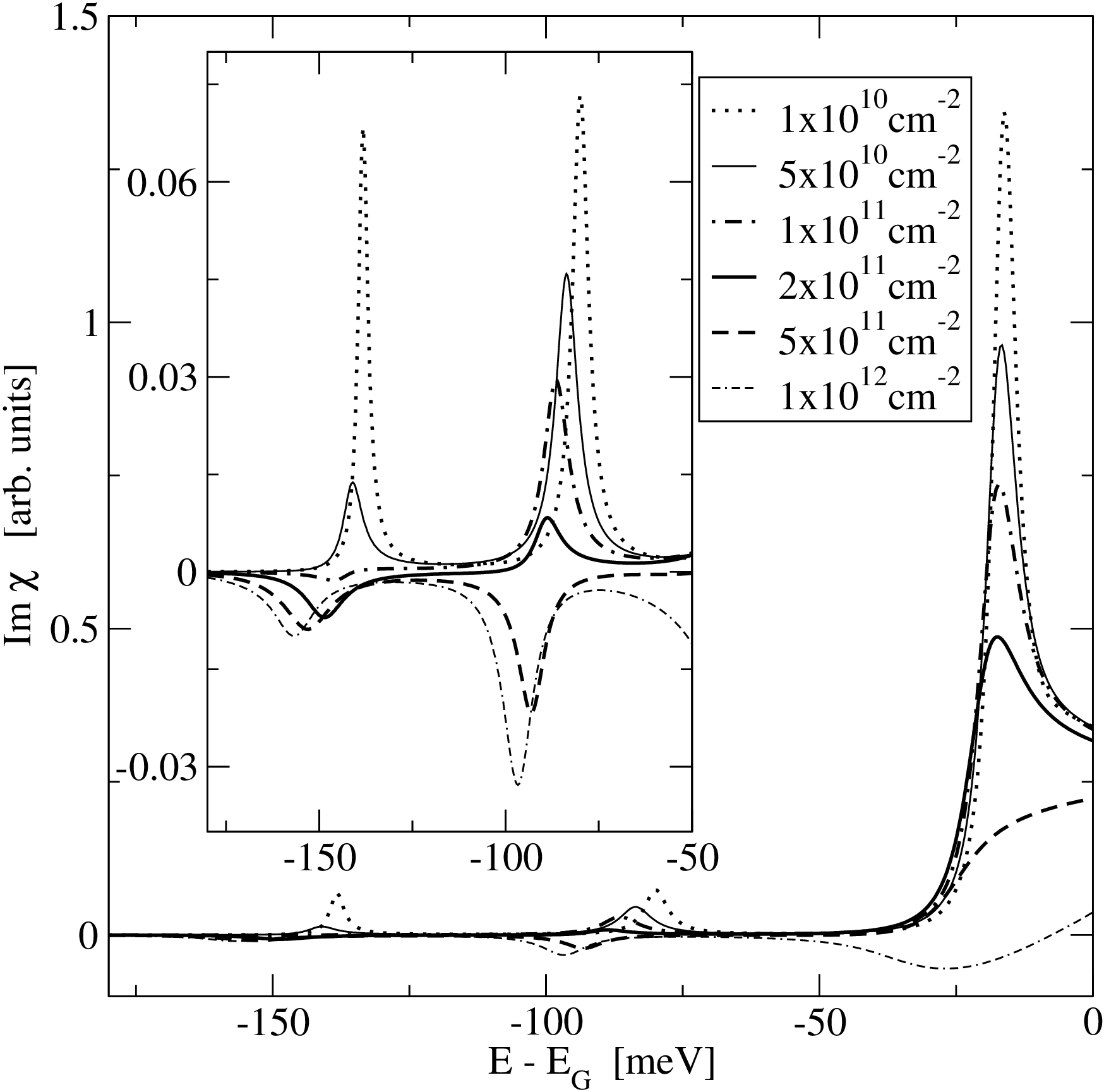}	
    \caption{Imaginary part of the optical susceptibility for the combined QD-WL 
    		system including interaction-induced dephasing and line shifts due to Coulomb interaction and
		carrier-phonon interaction
		for various total carrier densities. The inset shows a 
	 	scale up of the QD resonances.
	 \label{fig:both_dens}
	 }
    \end{center}
\end{figure}

In Fig.~\ref{fig:both_dens} calculated absorption spectra are shown, which include correlations due to 
carrier-carrier scattering Eqs.~\eqref{eq:sddcoul}-\eqref{eq:sodcoul}, and interaction with LO-phonons,
Eqs.~\eqref{eq:sddphon}-\eqref{eq:sodphon}, both evaluated with self-consistently renormalized single-particle 
energies.
As in the result for dephasing due to Coulomb interaction, Fig.~\ref{fig:coul_dens}, we see the 
bleaching and red-shift of the resonances due to many-body interactions and also the 
saturation of the s-shell gain. 
Although we observe that the Coulomb interaction is the clearly dominant
dephasing mechanism for high carrier densities, we also infer from a comparison of Fig.~\ref{fig:both_dens} 
with Fig.~\ref{fig:coul_dens} that even in the gain regime the electron-phonon interaction gives rise to a clear
increase in the dephasing. Nevertheless  polaronic features are absent in the spectra, since the complicated
multi-peak structure of the spectral function is completely damped out due to Coulomb effects.

\begin{figure}[!ht]
\vspace*{0.5cm}
    \begin{center}
    \includegraphics[width=.45\textwidth]{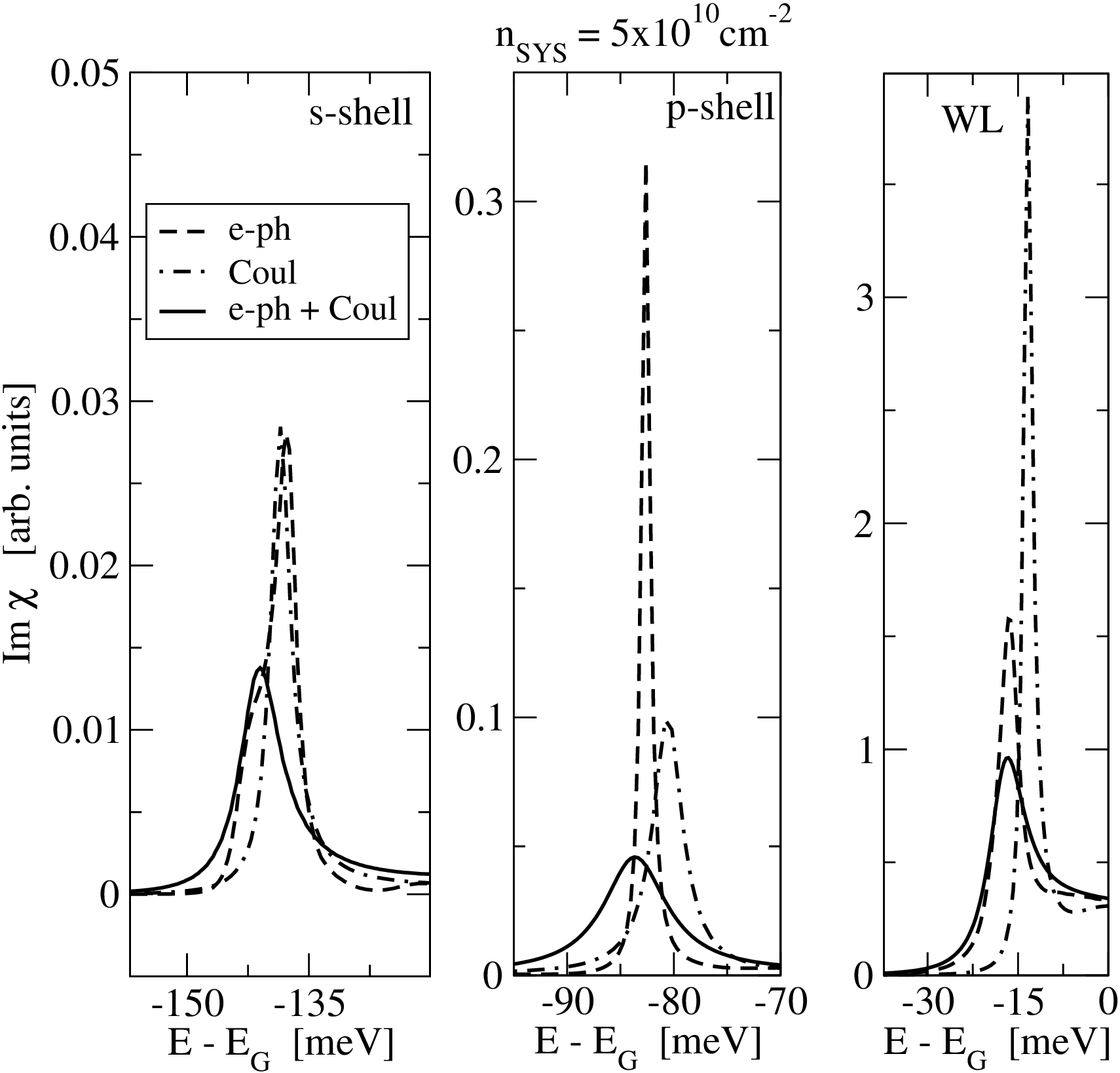}	
    \caption{Comparison of spectra with different dephasing mechanisms
    for a carrier density of $\dens{5}{10}$.
	 \label{fig:compare}
	 }
    \end{center}
\end{figure}

\begin{figure}[!ht]
\vspace*{0.5cm}
    \begin{center}
    \includegraphics[width=.45\textwidth]{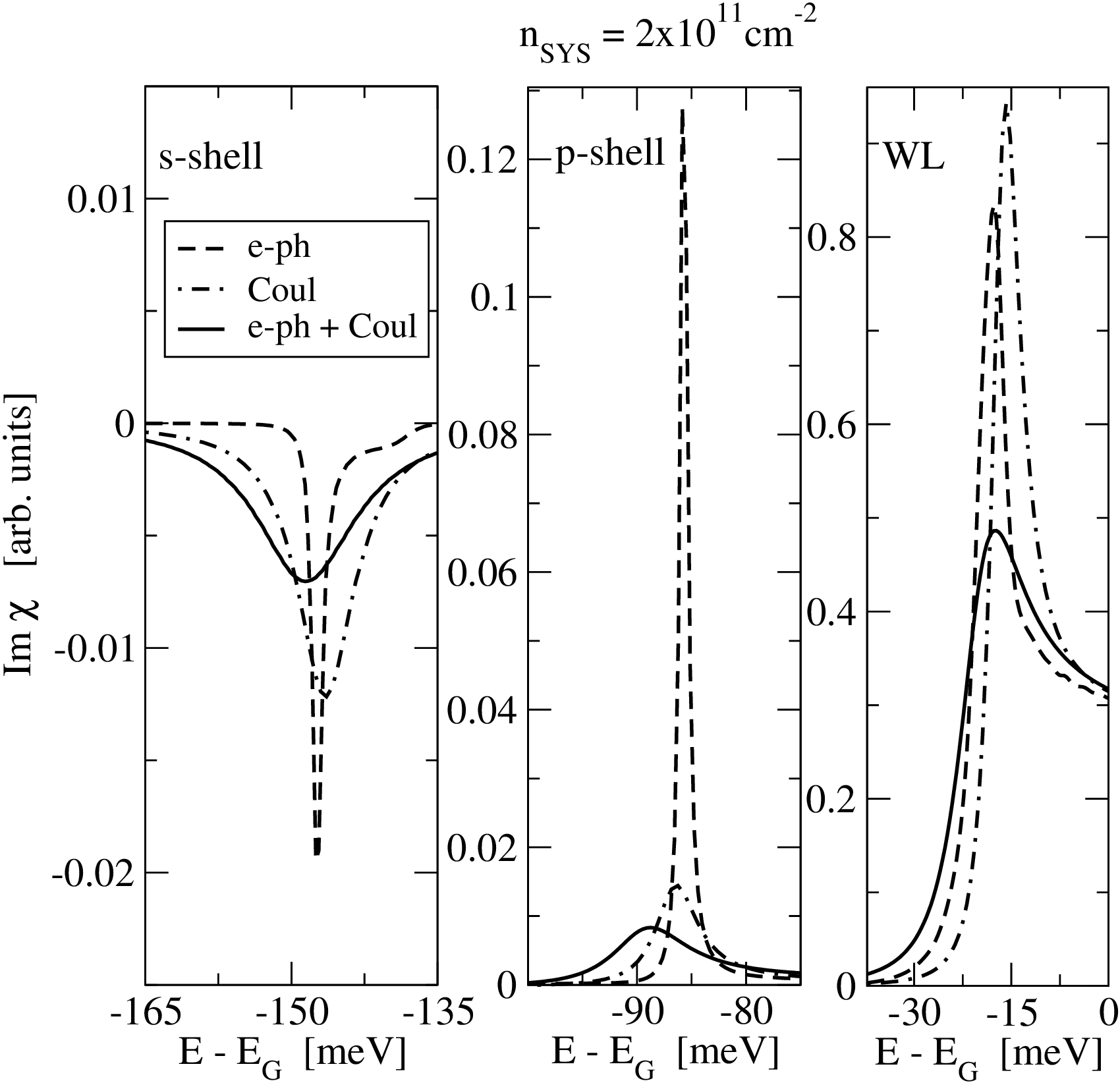}	
    \caption{Comparison of spectra with different dephasing mechanisms
    for a carrier density of $\dens{2}{11}$.
	 \label{fig:compare2}
	 }
    \end{center}
\end{figure}

For intermediate carrier densities around $\dens{5}{10}$ both types of 
interaction processes are equally important.
Comparing the results in Fig. \ref{fig:compare} one can conclude for our situation
that taking only the Coulomb dephasing mechanism into account underestimates the 
dephasing of the ground state transition by roughly a factor of two, while the damping of the WL is even
dominated by carrier-phonon interaction.
For higher carrier densities, however, this picture changes as can be seen in Fig.~\ref{fig:compare2}. 
Using a carrier density of $\dens{2}{11}$,
the Coulomb interaction is clearly the dominant mechanism for the QD resonances.
For the excitonic resonance of the WL, the two mechanisms are equally important even at this rather high carrier 
density, where we are already in the gain regime for the s-shell transition.

\subsubsection*{Comparison with Experiments}\label{sec:comexp}
In most experiments, the emission from an ensemble of QDs is studied such that the  
additional inhomogeneous broadening of the QD resonances due to size and composition 
fluctuations contributes. 
However, recent experiments \cite{Bayer:02,Matsuda:03} have been performed at room 
temperature on single QDs. In Fig. 1(a) of Ref.~\cite{Matsuda:03} 
optical spectra are displayed for different excitation intensities.
Although a direct comparison of the excitation dependence is not possible,
due to uncertainties in the carrier densities generated in those experiments, the general features 
of the spectra are identical.
A clear distinction between the QD transitions and the excitonic resonance 
of the WL is seen. 
The QD resonances show a density dependent bleaching and a pronounced red shift 
due to many-body correlations while the spectral position of the WL is almost unchanged and 
the WL resonance is only bleached out. This means that the red-shift of the QD resonances cannot be 
attributed to band-gap shrinkage effects of the WL. 

The observed homogeneous linewidth in this experiment is 8-13meV
depending on the excitation intensity. This is larger than our
findings, but one has to take into account that the QD, investigated
in Ref.~\cite{Matsuda:03}, seems to have three confined electronic
shells which results in more dephasing channels. A better comparison
is possible with the results of Ref.~\cite{Bayer:02} because two types
of QDs are investigated, where one type resembles our model in the
respect that it is supposed to have two confined electronic shells.
Regarding the s-shell, we can infer from Fig. 3 of
Ref.~\cite{Bayer:02} that the observed homogeneous linewidth is
slightly above 3meV at room temperature which is similar to our result
of 3.2meV. The calculated value is practically unchanged for a carrier
density range from $5\times 10^{8}$ to $10^{10}\,\text{cm}^{-2}$.

\begin{figure}[!ht]
\vspace*{0.5cm}
    \begin{center}
    \includegraphics[width=.4\textwidth]{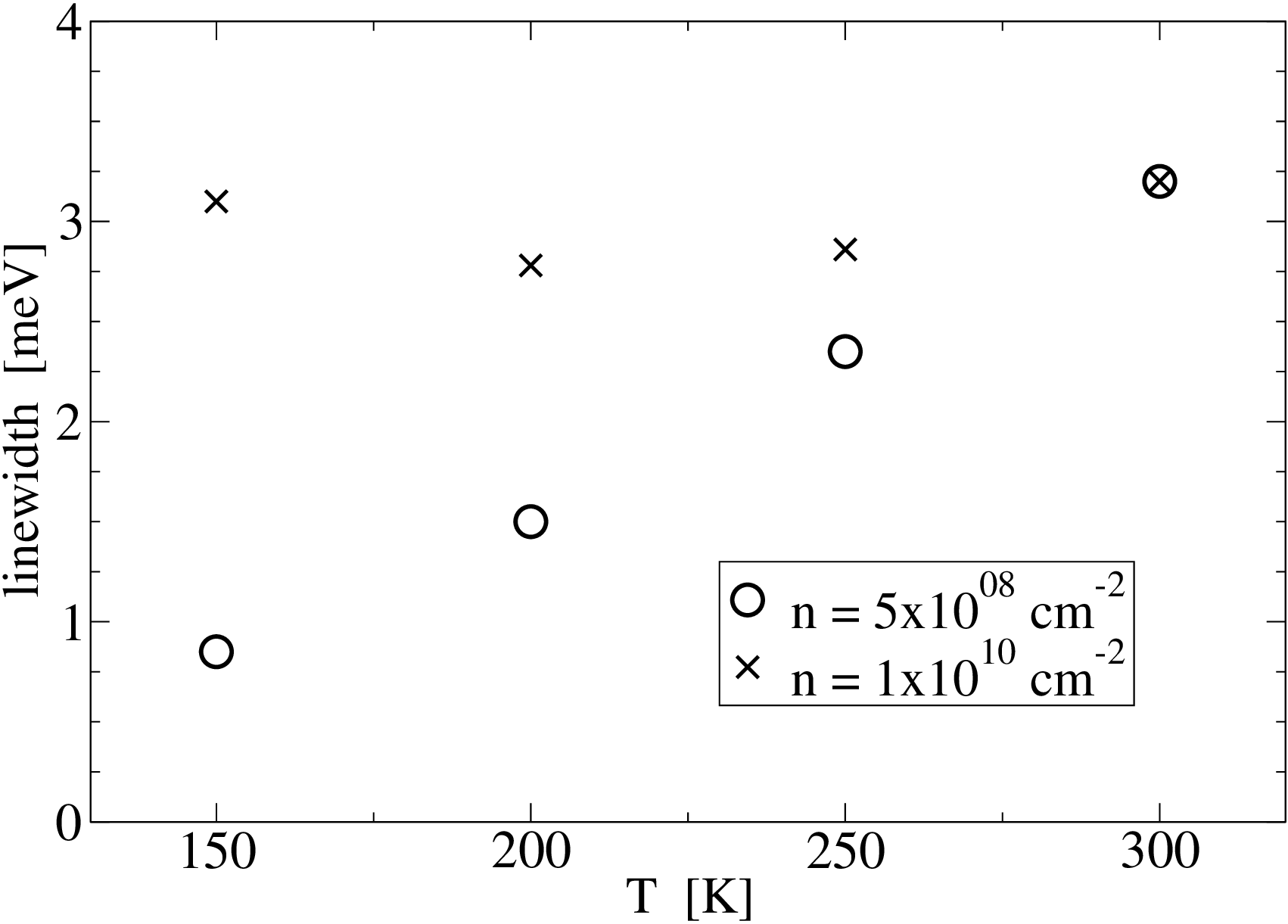}	
    \caption{Temperature dependence of the linewidth (full half-width) for the QD ground state resonance
and different carrier densities. 
	 \label{fig:temp_dep}
	 }
    \end{center}
\end{figure}

For a better comparison with experiments, we have also calculated the
temperature dependence of the homogeneous linewidth in a temperature
range in which the influence of LA-phonons remains small.  As shown in
Fig.~\ref{fig:temp_dep}, we obtain for low carrier densities (where
carrier-carrier scattering is sufficiently weak) the expected
reduction of dephasing with decreasing temperature as seen in
Ref.~\cite{Bayer:02}. (A quantitative comparison would require a
better knowledge of QD parameters and phonon modes in the QD-WL
system.)  For the case of a higher excitation density of
$10^{10}\,\text{cm}^{-2}$, the line-broadening is almost temperature
independent due to a balancing of different scattering channels. The
dephasing due to LO-phonons decreases with temperature as in the
low-density case. However, the Coulomb interaction is no longer
negligible for this carrier density. With decreasing temperature, the
WL states are less populated while the occupation of QD states
increases.  This enhances intra-dot relaxation processes and provides
stronger dephasing due to Coulomb interaction.

\section{Conclusion}\label{sec:concl}
We have calculated the excitation induced dephasing and lineshifts for a
QD-WL system on a microscopic basis. Both
Coulomb and LO-phonon contributions to the homogeneous linewidths are found to be equally important
for elevated carrier densities.
The role of self-consistent single-particle energy renormalizations in the scattering integrals
is emphasized.
For the Coulomb interaction the relative importance of various scattering channels
has been analyzed.


\begin{acknowledgments}
This work was supported by the Deutsche Forschungsgemeinschaft and
with a grant for CPU time at the NIC, Forschungszentrum J\"{u}lich.
\end{acknowledgments}

\begin{appendix}
\section{Quantum dot model system and Coulomb matrix elements}\label{sec:qdmodel}
We focus on lens-shaped InGaAs QDs which are located on top of a two dimensional WL.
Such systems typically consist of localized QD states energetically below a quasi-continuum of delocalized 
WL states for electrons and holes. 
In the present paper we assume the single particle spectrum as depicted 
in Fig. 1 of Ref.~\cite{Nielsen:04}. 
For the in plane motion we assume a parabolic confinement potential, leading to harmonic oscillator 
like single particle wavefunctions. 
To account for the finite depth of the confinement potential, 
we take into account two confined shells for electrons and holes, which we refer to as s-shell 
for the ground state and p-shell for the double degenerated excited state, due to
their angular momentum properties. 
We choose a level spacing of 40meV for electrons and 15meV for holes
so that the free carrier transitions appear at -55meV relative to the WL bandedge 
for the p-shell and at -110meV for the s-shell.
The WL states are modeled by orthogonalized plane waves (OPW) \cite{Nielsen:04}. 
For the motion in growth direction (z-direction) we assume a 
infinite height potential well of 4 nm thickness. 

Under this assumptions we can construct the Coulomb matrix elements
\begin{eqnarray}\label{eq:coulmatr}
    V_{\alpha\beta\gamma\delta} &=& \frac{1}{A} \: \sum \limits_{\bf q} \, 
       V_{\bf q} \, 
\nonumber \\
    && \times \int d^2 \varrho \: \varphi^\ast_\alpha({\bm \varrho})
       \varphi_{\delta}({\bm \varrho})  \: e^{-i {\bf q} \cdot
       {\bm \varrho}} \;
\nonumber \\
    && \times   \int d^2 \varrho' \: \varphi^\ast_{\beta}({\bm \varrho}')
       \varphi_{\gamma}({\bm \varrho}') \: e^{i {\bf q} \cdot
       {\bm \varrho}'}
\label{eq:Coulomb2} ~,
\end{eqnarray}
consisting of overlap integrals between single-particle wavefunctions and the Coulomb potential $V_{\bf q}$
as given in Ref.~\cite{Nielsen:04}.
For the screening we use a generalization of the static Lindhard formula
which is also explained in detail in Ref.~\cite{Nielsen:04}. This procedure leads to the replacement
$V_{\bf q}\rightarrow W_{\bf q}$ in Eq.~\eqref{eq:coulmatr} for the matrix elements $W_{\alpha\beta\gamma\delta}$.
Because we are working in the envelope
approximation with equal envelope wavefunctions, we do not need to 
consider band indices for the Coulomb matrix elements.
All other parameters are chosen as in Ref.~\cite{Nielsen:04}.


\section{Retarded polaronic Green's function}\label{sec:polGF}

The retarded polaronic GFs obeys the equation of motion
\begin{equation}
\begin{split}
   \Big[ i\hbar\pdiff{}{\tau} - e^a_{\alpha} \Big]&
     ~G^{\text{a,ret}}_{\alpha}(\tau)
   =  \delta(\tau) \\
   &+ \int\!d\tau' ~~ \Sigma^{\text{a,ret}}_{\alpha}(\tau-\tau') ~ G^{\text{a,ret}}_{\alpha}(\tau')~.
\label{eq:Dyson_G_ret}
\end{split}
\end{equation}
The corresponding retarded self-energy is given in RPA by
\begin{equation}
      \Sigma^{\text{a,ret}}_{\alpha}(\tau) = i \hbar \sum_{\beta} ~
      G^{\text{a,ret}}_{\beta}(\tau) ~ D^<_{\beta \alpha}(-\tau).
\label{eq:Sigma_ret}
\end{equation}
We assume for the calculations restricted to electron-phonon interaction
that the polaronic retarded GF is not influenced by 
population effects. This has been verified
for a bulk system in Ref.~\cite{Gartner:99}.

Assuming that the phonon system is in thermal equilibrium,
the phonon propagator (combined with the interaction matrix elements) 
is given by
\begin{equation}
\begin{split}
  i\hbar ~ D^<_{\beta \alpha}(\tau)&=   \frac{M^2_{LO}}{e^2/\varepsilon_0} V_{\beta\alpha\beta\alpha}\\ \times
   &\Big[    N_\text{LO}  ~ e^{-i\omega_\text{LO}\tau}
      ~+~(1+N_\text{LO}) ~ e^{ i\omega_\text{LO}\tau} \Big]
\label{eq:pn_propagator}
\end{split}
\end{equation}
where we consider monochromatic LO-phonons with the frequency $\omega_\text{LO}$. 
The corresponding population of the phonon bath is given by a Bose-Einstein distribution
$n_\text{LO}=1/(e^{\hbar\omega_\text{LO}/kT}-1)$ and 
$V_{\beta\alpha\beta\alpha}$ is the unscreened 
Coulomb matrix element (cf. section \ref{sec:qdmodel}).


\section{Connection between memory effects and frequency dependence in the scattering integrals}\label{sec:memory}
In this section we is show how the non-Markovian (Markovian) scattering integrals
lead to frequency dependent (independent) dephasing terms.
The time domain formulation of the SBE with correlation contributions due to Coulomb interaction is 
given by \cite{Haug:94}
\begin{equation}
\begin{split}
 (i\hbar\frac{\partial}{\partial t} 
-\varepsilon^{e,\text{HF}}_\alpha-\varepsilon^{h,\text{HF}}_\alpha)&\psi_{\alpha}(t)
+\left[1-f^e_{\alpha}- f^h_{\alpha} \right ]\Omega^\text{HF}_\alpha(t)\\
 =&\ -i\hbar\ S^\text{Coul}_\alpha(t)~.
\end{split}
\end{equation}
In the following we explicitly consider the direct e-e and h-h interaction contributions of the 
diagonal dephasing,
\begin{widetext}
\begin{eqnarray}
S_\alpha(t)&=&\hbar^2\int\limits_{-\infty}^t dt' \sum\limits_{\substack{
a,b=e,h\\
b\neq a}}\, 
\sum\limits_{\alpha_1\alpha_2\alpha_3}
2W_{\alpha\alpha_2\alpha_3\alpha_1}(t)
W^\ast_{\alpha\alpha_2\alpha_3\alpha_1}(t')\\
&\times&\nonumber
G^\text{b,ret}_{\alpha}(t,t')
\left[G^\text{a,ret}_{\alpha_2}(t,t')\right]^\ast
G^\text{a,ret}_{\alpha_3}(t,t')
G^\text{a,ret}_{\alpha_1}(t,t')\
\psi_\alpha(t')F_{\alpha_2\alpha_3\alpha_1}(t')~.
\end{eqnarray}
\end{widetext}
Other terms can be treated in complete analogy.
Here we have defined
\begin{eqnarray}
F_{\alpha_2\alpha_3\alpha_1}(t)&=&\nonumber
\left(1-f^a_{\alpha_2}(t)\right)f^a_{\alpha_3}(t)f^a_{\alpha_1}(t)\\
&&+(f\rightarrow 1-f)~.
\end{eqnarray}
We assume static screening where the Coulomb matrix elements depend only 
parametrically on time via the population functions. For the discussed excitation
conditions both $f$ and $W$ are time independent.
With the ansatz 
\begin{eqnarray}
G^\text{a,ret}_{\alpha}(t,t')=-\frac{i}{\hbar}\Theta(t-t')e^{-\frac{i}{\hbar}\tilde\varepsilon^a_\alpha(t-t')}~,
\end{eqnarray} 
which corresponds to Eq.~\eqref{eq:renGret},
we obtain
\begin{eqnarray}
S_\alpha(t)&=&\frac{1}{\hbar^2}\sum\limits_{\substack{
a,b=e,h\\
b\neq a}}\, \nonumber
\sum\limits_{\alpha_1\alpha_2\alpha_3}
2W_{\alpha\alpha_2\alpha_3\alpha_1}
W^\ast_{\alpha\alpha_2\alpha_3\alpha_1}
F_{\alpha_2\alpha_3\alpha_1}\\
&\times&\nonumber
\int\limits_{-\infty}^t dt'
e^{-\frac{i}{\hbar} (\tilde\varepsilon^{b}_\alpha+\tilde\varepsilon^a_{\alpha_1}-
\left(\tilde\varepsilon^a_{\alpha_2}\right)^\ast+\tilde\varepsilon^a_{\alpha_3})(t-t')}
\psi_\alpha(t')~.
\end{eqnarray} 
With the Fourier transform
$\psi_\alpha(t)=\int \frac{d\omega}{2\pi}  e^{-i\omega t}\ \psi_\alpha(\omega)$
and using the integral relation
\begin{equation}\label{eq:intRel}
\int\limits_{-\infty}^t dt' e^{\frac{i}{\hbar} \Delta (t-t')}=\frac{i \hbar}{\Delta},\quad \text{Im}\Delta>0
\end{equation}
we arrive at Eq.~\eqref{eq:sbe} and the terms of Eq.~\eqref{eq:sddcoul} which are discussed here.

For the Markov approximation one separates from the transition amplitude
a phase factor, which rapidly oscillates with the (renormalized) interband transition energy,
via the ansatz
\begin{equation}
\psi_\alpha(t)=
e^{\frac{i}{\hbar}(
\tilde\varepsilon^e_{\alpha}+
\tilde\varepsilon^h_{\alpha})t}
\tilde\psi_\alpha(t)~.
\end{equation}
Now one assumes that a weakly time dependent $\tilde\psi_\alpha(t')$ 
can be replaced by the transition amplitude at the actual time $t$, 
$\tilde\psi_\alpha(t)$, thus neglecting 
the memory effects of $\tilde\psi_\alpha$. 
Using the integral relation \eqref{eq:intRel} again, 
one arrives at a expression similar to Eq.~\eqref{eq:sddcoul} but with
$g\left(\hbar\omega-\tilde\varepsilon^{b}_\alpha-\tilde\varepsilon^a_{\alpha_1}+
\left(\tilde\varepsilon^a_{\alpha_2}\right)^\ast-\tilde\varepsilon^a_{\alpha_3}\right)$
replaced by $g\left(\tilde\varepsilon^{a}_\alpha-\tilde\varepsilon^a_{\alpha_1}+
\left(\tilde\varepsilon^a_{\alpha_2}\right)^\ast-\tilde\varepsilon^a_{\alpha_3}\right)$.

\end{appendix}


\begin{thebibliography}{28}
\expandafter\ifx\csname natexlab\endcsname\relax\def\natexlab#1{#1}\fi
\expandafter\ifx\csname bibnamefont\endcsname\relax
  \def\bibnamefont#1{#1}\fi
\expandafter\ifx\csname bibfnamefont\endcsname\relax
  \def\bibfnamefont#1{#1}\fi
\expandafter\ifx\csname citenamefont\endcsname\relax
  \def\citenamefont#1{#1}\fi
\expandafter\ifx\csname url\endcsname\relax
  \def\url#1{\texttt{#1}}\fi
\expandafter\ifx\csname urlprefix\endcsname\relax\def\urlprefix{URL }\fi
\providecommand{\bibinfo}[2]{#2}
\providecommand{\eprint}[2][]{\url{#2}}

\bibitem[{\citenamefont{Masumoto and Takagahara}(2002)}]{Masumoto:02}
\bibinfo{editor}{\bibfnamefont{Y.}~\bibnamefont{Masumoto}} \bibnamefont{and}
  \bibinfo{editor}{\bibfnamefont{T.}~\bibnamefont{Takagahara}}, eds.,
  \emph{\bibinfo{title}{Semiconductor Quantum Dots}}
  (\bibinfo{publisher}{Springer-Verlag}, \bibinfo{address}{Berlin},
  \bibinfo{year}{2002}), \bibinfo{edition}{1st} ed.

\bibitem[{\citenamefont{Michler}(2003)}]{Michler:03}
\bibinfo{editor}{\bibfnamefont{P.}~\bibnamefont{Michler}}, ed.,
  \emph{\bibinfo{title}{Single Quantum Dots}}
  (\bibinfo{publisher}{Springer-Verlag}, \bibinfo{address}{Berlin},
  \bibinfo{year}{2003}), \bibinfo{edition}{1st} ed.

\bibitem[{\citenamefont{Michler et~al.}(2000)\citenamefont{Michler, Imamoglu,
  Mason, Carson, Strouse, and Buratto}}]{Michler:00}
\bibinfo{author}{\bibfnamefont{P.}~\bibnamefont{Michler}},
  \bibinfo{author}{\bibfnamefont{A.}~\bibnamefont{Imamoglu}},
  \bibinfo{author}{\bibfnamefont{M.~D.} \bibnamefont{Mason}},
  \bibinfo{author}{\bibfnamefont{P.~J.} \bibnamefont{Carson}},
  \bibinfo{author}{\bibfnamefont{G.~F.} \bibnamefont{Strouse}},
  \bibnamefont{and} \bibinfo{author}{\bibfnamefont{S.~K.}
  \bibnamefont{Buratto}}, \bibinfo{journal}{Nature}
  \textbf{\bibinfo{volume}{406}}, \bibinfo{pages}{968} (\bibinfo{year}{2000}).

\bibitem[{\citenamefont{Moreau et~al.}(2001)\citenamefont{Moreau, Robert,
  Manin, Thierry-Mieg, Gerad, and Abram}}]{Moreau:01}
\bibinfo{author}{\bibfnamefont{E.}~\bibnamefont{Moreau}},
  \bibinfo{author}{\bibfnamefont{I.}~\bibnamefont{Robert}},
  \bibinfo{author}{\bibfnamefont{L.}~\bibnamefont{Manin}},
  \bibinfo{author}{\bibfnamefont{V.}~\bibnamefont{Thierry-Mieg}},
  \bibinfo{author}{\bibfnamefont{J.~M.} \bibnamefont{Gerad}}, \bibnamefont{and}
  \bibinfo{author}{\bibfnamefont{I.}~\bibnamefont{Abram}},
  \bibinfo{journal}{Phys. Rev. Lett.} \textbf{\bibinfo{volume}{87}},
  \bibinfo{pages}{183601} (\bibinfo{year}{2001}).

\bibitem[{\citenamefont{Pelton et~al.}(2002)\citenamefont{Pelton, Santori,
  Vuckovic, Zhang, Solomon, Plant, and Yamamoto}}]{Pelton:02}
\bibinfo{author}{\bibfnamefont{M.}~\bibnamefont{Pelton}},
  \bibinfo{author}{\bibfnamefont{C.}~\bibnamefont{Santori}},
  \bibinfo{author}{\bibfnamefont{J.}~\bibnamefont{Vuckovic}},
  \bibinfo{author}{\bibfnamefont{B.}~\bibnamefont{Zhang}},
  \bibinfo{author}{\bibfnamefont{G.~S.} \bibnamefont{Solomon}},
  \bibinfo{author}{\bibfnamefont{J.}~\bibnamefont{Plant}}, \bibnamefont{and}
  \bibinfo{author}{\bibfnamefont{Y.}~\bibnamefont{Yamamoto}},
  \bibinfo{journal}{Phys. Rev. Lett.} \textbf{\bibinfo{volume}{89}},
  \bibinfo{pages}{233602} (\bibinfo{year}{2002}).

\bibitem[{\citenamefont{Reithmaier et~al.}(2004)\citenamefont{Reithmaier, Sek,
  L{\"o}ffler, Hofmann, Kuhn, Reitzenstein, Keldysh, Kulakovskii, Reinecke, and
  Forchel}}]{Reithmaier:04}
\bibinfo{author}{\bibfnamefont{J.~P.} \bibnamefont{Reithmaier}},
  \bibinfo{author}{\bibfnamefont{G.}~\bibnamefont{Sek}},
  \bibinfo{author}{\bibfnamefont{A.}~\bibnamefont{L{\"o}ffler}},
  \bibinfo{author}{\bibfnamefont{C.}~\bibnamefont{Hofmann}},
  \bibinfo{author}{\bibfnamefont{S.}~\bibnamefont{Kuhn}},
  \bibinfo{author}{\bibfnamefont{S.}~\bibnamefont{Reitzenstein}},
  \bibinfo{author}{\bibfnamefont{L.~V.} \bibnamefont{Keldysh}},
  \bibinfo{author}{\bibfnamefont{V.~D.} \bibnamefont{Kulakovskii}},
  \bibinfo{author}{\bibfnamefont{T.~L.} \bibnamefont{Reinecke}},
  \bibnamefont{and} \bibinfo{author}{\bibfnamefont{A.}~\bibnamefont{Forchel}},
  \bibinfo{journal}{Nature} \textbf{\bibinfo{volume}{\textbf{432}}},
  \bibinfo{pages}{197} (\bibinfo{year}{2004}).

\bibitem[{\citenamefont{Yoshie et~al.}(2004)\citenamefont{Yoshie, Scherer,
  Hendrickson, Khitrova, Gibbs, Rupper, Ell, Shchekin, and Deppe}}]{Yoshie:04}
\bibinfo{author}{\bibfnamefont{T.}~\bibnamefont{Yoshie}},
  \bibinfo{author}{\bibfnamefont{A.}~\bibnamefont{Scherer}},
  \bibinfo{author}{\bibfnamefont{J.}~\bibnamefont{Hendrickson}},
  \bibinfo{author}{\bibfnamefont{G.}~\bibnamefont{Khitrova}},
  \bibinfo{author}{\bibfnamefont{H.~M.} \bibnamefont{Gibbs}},
  \bibinfo{author}{\bibfnamefont{G.}~\bibnamefont{Rupper}},
  \bibinfo{author}{\bibfnamefont{C.}~\bibnamefont{Ell}},
  \bibinfo{author}{\bibfnamefont{O.~B.} \bibnamefont{Shchekin}},
  \bibnamefont{and} \bibinfo{author}{\bibfnamefont{D.~G.} \bibnamefont{Deppe}},
  \bibinfo{journal}{Nature} \textbf{\bibinfo{volume}{\textbf{432}}},
  \bibinfo{pages}{200} (\bibinfo{year}{2004}).

\bibitem[{\citenamefont{Hu et~al.}(1996)\citenamefont{Hu, Gie{\ss}en,
  Peyghambarian, and Koch}}]{Hu:96}
\bibinfo{author}{\bibfnamefont{Y.~Z.} \bibnamefont{Hu}},
  \bibinfo{author}{\bibfnamefont{H.}~\bibnamefont{Gie{\ss}en}},
  \bibinfo{author}{\bibfnamefont{N.}~\bibnamefont{Peyghambarian}},
  \bibnamefont{and} \bibinfo{author}{\bibfnamefont{S.~W.} \bibnamefont{Koch}},
  \bibinfo{journal}{Phys. Rev. B} \textbf{\bibinfo{volume}{\textbf{53}}},
  \bibinfo{pages}{4814} (\bibinfo{year}{1996}).

\bibitem[{\citenamefont{Schneider et~al.}(2001)\citenamefont{Schneider, Chow,
  and Koch}}]{Schneider:01a}
\bibinfo{author}{\bibfnamefont{H.~C.} \bibnamefont{Schneider}},
  \bibinfo{author}{\bibfnamefont{W.~W.} \bibnamefont{Chow}}, \bibnamefont{and}
  \bibinfo{author}{\bibfnamefont{S.~W.} \bibnamefont{Koch}},
  \bibinfo{journal}{Phys. Rev. B} \textbf{\bibinfo{volume}{\textbf{64}}},
  \bibinfo{pages}{115315} (\bibinfo{year}{2001}).

\bibitem[{\citenamefont{Schneider et~al.}(2004)\citenamefont{Schneider, Chow,
  and Koch}}]{Schneider:04}
\bibinfo{author}{\bibfnamefont{H.~C.} \bibnamefont{Schneider}},
  \bibinfo{author}{\bibfnamefont{W.~W.} \bibnamefont{Chow}}, \bibnamefont{and}
  \bibinfo{author}{\bibfnamefont{S.~W.} \bibnamefont{Koch}},
  \bibinfo{journal}{Phys. Rev. B} \textbf{\bibinfo{volume}{\textbf{70}}},
  \bibinfo{pages}{235308} (\bibinfo{year}{2004}).

\bibitem[{\citenamefont{Manzke and Henneberger}(2002)}]{Manzke:02}
\bibinfo{author}{\bibfnamefont{G.}~\bibnamefont{Manzke}} \bibnamefont{and}
  \bibinfo{author}{\bibfnamefont{K.}~\bibnamefont{Henneberger}},
  \bibinfo{journal}{phys. stat. sol. (b)}
  \textbf{\bibinfo{volume}{\textbf{234}}}, \bibinfo{pages}{233}
  (\bibinfo{year}{2002}).

\bibitem[{\citenamefont{Uskov et~al.}(2000)\citenamefont{Uskov, Jauho,
  Tromborg, M{\o}rk, and Lang}}]{Uskov:00}
\bibinfo{author}{\bibfnamefont{A.~V.} \bibnamefont{Uskov}},
  \bibinfo{author}{\bibfnamefont{A.-P.} \bibnamefont{Jauho}},
  \bibinfo{author}{\bibfnamefont{B.}~\bibnamefont{Tromborg}},
  \bibinfo{author}{\bibfnamefont{J.}~\bibnamefont{M{\o}rk}}, \bibnamefont{and}
  \bibinfo{author}{\bibfnamefont{R.}~\bibnamefont{Lang}},
  \bibinfo{journal}{Phys. Rev. Lett.} \textbf{\bibinfo{volume}{\textbf{85}}},
  \bibinfo{pages}{1516} (\bibinfo{year}{2000}).

\bibitem[{\citenamefont{Krummheuer et~al.}(2002)\citenamefont{Krummheuer, Axt,
  and Kuhn}}]{Krummheuer:02}
\bibinfo{author}{\bibfnamefont{B.}~\bibnamefont{Krummheuer}},
  \bibinfo{author}{\bibfnamefont{V.~M.} \bibnamefont{Axt}}, \bibnamefont{and}
  \bibinfo{author}{\bibfnamefont{T.}~\bibnamefont{Kuhn}},
  \bibinfo{journal}{Phys. Rev. B} \textbf{\bibinfo{volume}{\textbf{65}}},
  \bibinfo{pages}{195313} (\bibinfo{year}{2002}).

\bibitem[{\citenamefont{Muljarov and Zimmermann}(2000)}]{Muljarov:04}
\bibinfo{author}{\bibfnamefont{E.~A.} \bibnamefont{Muljarov}} \bibnamefont{and}
  \bibinfo{author}{\bibfnamefont{R.}~\bibnamefont{Zimmermann}},
  \bibinfo{journal}{Phys. Rev. Lett.} \textbf{\bibinfo{volume}{\textbf{85}}},
  \bibinfo{pages}{1516} (\bibinfo{year}{2000}).

\bibitem[{\citenamefont{Borri et~al.}(2001)\citenamefont{Borri, Langbein,
  Schneider, Woggon, Sellin, Ouyang, and Bimberg}}]{Borri:01}
\bibinfo{author}{\bibfnamefont{P.}~\bibnamefont{Borri}},
  \bibinfo{author}{\bibfnamefont{W.}~\bibnamefont{Langbein}},
  \bibinfo{author}{\bibfnamefont{S.}~\bibnamefont{Schneider}},
  \bibinfo{author}{\bibfnamefont{U.}~\bibnamefont{Woggon}},
  \bibinfo{author}{\bibfnamefont{R.~L.} \bibnamefont{Sellin}},
  \bibinfo{author}{\bibfnamefont{D.}~\bibnamefont{Ouyang}}, \bibnamefont{and}
  \bibinfo{author}{\bibfnamefont{D.}~\bibnamefont{Bimberg}},
  \bibinfo{journal}{Phys. Rev. Lett.} \textbf{\bibinfo{volume}{\textbf{87}}},
  \bibinfo{pages}{157401} (\bibinfo{year}{2001}).

\bibitem[{\citenamefont{Htoon et~al.}(2001)\citenamefont{Htoon, Kulik,
  Baklenov, Jr., Takagahara, and Shih}}]{Htoon:01}
\bibinfo{author}{\bibfnamefont{H.}~\bibnamefont{Htoon}},
  \bibinfo{author}{\bibfnamefont{D.}~\bibnamefont{Kulik}},
  \bibinfo{author}{\bibfnamefont{O.}~\bibnamefont{Baklenov}},
  \bibinfo{author}{\bibfnamefont{A.~L.~H.} \bibnamefont{Jr.}},
  \bibinfo{author}{\bibfnamefont{T.}~\bibnamefont{Takagahara}},
  \bibnamefont{and} \bibinfo{author}{\bibfnamefont{C.~K.} \bibnamefont{Shih}},
  \bibinfo{journal}{Phys. Rev. B} \textbf{\bibinfo{volume}{\textbf{63}}},
  \bibinfo{pages}{241303(R)} (\bibinfo{year}{2001}).

\bibitem[{\citenamefont{Oulton et~al.}(2003)\citenamefont{Oulton, Finley,
  Tartakovskii, Mowbray, Skolnick, Hopkinson, Vasanelli, Ferreira, and
  Bastard}}]{Oulton:03}
\bibinfo{author}{\bibfnamefont{R.}~\bibnamefont{Oulton}},
  \bibinfo{author}{\bibfnamefont{J.~J.} \bibnamefont{Finley}},
  \bibinfo{author}{\bibfnamefont{A.~I.} \bibnamefont{Tartakovskii}},
  \bibinfo{author}{\bibfnamefont{D.~J.} \bibnamefont{Mowbray}},
  \bibinfo{author}{\bibfnamefont{M.~S.} \bibnamefont{Skolnick}},
  \bibinfo{author}{\bibfnamefont{M.}~\bibnamefont{Hopkinson}},
  \bibinfo{author}{\bibfnamefont{A.}~\bibnamefont{Vasanelli}},
  \bibinfo{author}{\bibfnamefont{R.}~\bibnamefont{Ferreira}}, \bibnamefont{and}
  \bibinfo{author}{\bibfnamefont{G.}~\bibnamefont{Bastard}},
  \bibinfo{journal}{Phys. Rev. B} \textbf{\bibinfo{volume}{\textbf{68}}},
  \bibinfo{pages}{235301} (\bibinfo{year}{2003}).

\bibitem[{\citenamefont{Inoshita and Sakaki}(1997)}]{Inoshita:97}
\bibinfo{author}{\bibfnamefont{T.}~\bibnamefont{Inoshita}} \bibnamefont{and}
  \bibinfo{author}{\bibfnamefont{H.}~\bibnamefont{Sakaki}},
  \bibinfo{journal}{Phys. Rev. B} \textbf{\bibinfo{volume}{\textbf{56}}},
  \bibinfo{pages}{4355} (\bibinfo{year}{1997}).

\bibitem[{\citenamefont{Seebeck et~al.}(2005)\citenamefont{Seebeck, Nielsen,
  Gartner, and Jahnke}}]{Seebeck:05}
\bibinfo{author}{\bibfnamefont{J.}~\bibnamefont{Seebeck}},
  \bibinfo{author}{\bibfnamefont{T.~R.} \bibnamefont{Nielsen}},
  \bibinfo{author}{\bibfnamefont{P.}~\bibnamefont{Gartner}}, \bibnamefont{and}
  \bibinfo{author}{\bibfnamefont{F.}~\bibnamefont{Jahnke}},
  \bibinfo{journal}{Phys. Rev. B} \textbf{\bibinfo{volume}{\textbf{71}}},
  \bibinfo{pages}{125327} (\bibinfo{year}{2005}).

\bibitem[{\citenamefont{Bayer and Forchel}(2002)}]{Bayer:02}
\bibinfo{author}{\bibfnamefont{M.}~\bibnamefont{Bayer}} \bibnamefont{and}
  \bibinfo{author}{\bibfnamefont{A.}~\bibnamefont{Forchel}},
  \bibinfo{journal}{Phys. Rev. B} \textbf{\bibinfo{volume}{\textbf{65}}},
  \bibinfo{pages}{041308(R)} (\bibinfo{year}{2002}).

\bibitem[{\citenamefont{Matsuda et~al.}(2003)\citenamefont{Matsuda, Ikeda,
  Saiki, Saito, and Nishi}}]{Matsuda:03}
\bibinfo{author}{\bibfnamefont{K.}~\bibnamefont{Matsuda}},
  \bibinfo{author}{\bibfnamefont{K.}~\bibnamefont{Ikeda}},
  \bibinfo{author}{\bibfnamefont{T.}~\bibnamefont{Saiki}},
  \bibinfo{author}{\bibfnamefont{H.}~\bibnamefont{Saito}}, \bibnamefont{and}
  \bibinfo{author}{\bibfnamefont{K.}~\bibnamefont{Nishi}},
  \bibinfo{journal}{Appl. Phys. Lett.} \textbf{\bibinfo{volume}{\textbf{83}}},
  \bibinfo{pages}{2250} (\bibinfo{year}{2003}).

\bibitem[{\citenamefont{Haug and Koch}(1994)}]{Haug:94}
\bibinfo{author}{\bibfnamefont{H.}~\bibnamefont{Haug}} \bibnamefont{and}
  \bibinfo{author}{\bibfnamefont{S.~W.} \bibnamefont{Koch}},
  \emph{\bibinfo{title}{Quantum Theory of the Optical and Electronic Properties
  of Semiconductors}} (\bibinfo{publisher}{World Scientific Publ.},
  \bibinfo{address}{Singapore}, \bibinfo{year}{1994}), \bibinfo{edition}{3rd}
  ed.

\bibitem[{\citenamefont{Jahnke et~al.}(1997)\citenamefont{Jahnke, Kira, and
  Koch}}]{Jahnke:97}
\bibinfo{author}{\bibfnamefont{F.}~\bibnamefont{Jahnke}},
  \bibinfo{author}{\bibfnamefont{M.}~\bibnamefont{Kira}}, \bibnamefont{and}
  \bibinfo{author}{\bibfnamefont{S.~W.} \bibnamefont{Koch}},
  \bibinfo{journal}{Z. Physik B} \textbf{\bibinfo{volume}{\textbf{104}}},
  \bibinfo{pages}{559} (\bibinfo{year}{1997}).

\bibitem[{\citenamefont{Nielsen et~al.}(2005)\citenamefont{Nielsen, Gartner,
  Lorke, Seebeck, and Jahnke}}]{Nielsen:05}
\bibinfo{author}{\bibfnamefont{T.~R.} \bibnamefont{Nielsen}},
  \bibinfo{author}{\bibfnamefont{P.}~\bibnamefont{Gartner}},
  \bibinfo{author}{\bibfnamefont{M.}~\bibnamefont{Lorke}},
  \bibinfo{author}{\bibfnamefont{J.}~\bibnamefont{Seebeck}}, \bibnamefont{and}
  \bibinfo{author}{\bibfnamefont{F.}~\bibnamefont{Jahnke}},
  \bibinfo{journal}{Phys. Rev. B} \textbf{\bibinfo{volume}{\textbf{72}}},
  \bibinfo{pages}{235311} (\bibinfo{year}{2005}).




\bibitem[{\citenamefont{Lipavsk{\'y} et~al.}(1986)\citenamefont{Lipavsk{\'y},
  {\u{S}}pi{\u{c}}ka, and Velick{\'y}}}]{Lipa:86}
\bibinfo{author}{\bibfnamefont{P.}~\bibnamefont{Lipavsk{\'y}}},
  \bibinfo{author}{\bibfnamefont{V.}~\bibnamefont{{\u{S}}pi{\u{c}}ka}},
  \bibnamefont{and}
  \bibinfo{author}{\bibfnamefont{B.}~\bibnamefont{Velick{\'y}}},
  \bibinfo{journal}{Phys. Rev. B} \textbf{\bibinfo{volume}{\textbf{34}}},
  \bibinfo{pages}{6933} (\bibinfo{year}{1986}).

\bibitem[{\citenamefont{Tso and Morgenstern~Horing}(1991)}]{Tso:91}
\bibinfo{author}{\bibfnamefont{H.~C.} \bibnamefont{Tso}} \bibnamefont{and}
  \bibinfo{author}{\bibfnamefont{N.~J.} \bibnamefont{Morgenstern~Horing}},
  \bibinfo{journal}{Phys. Rev. B} \textbf{\bibinfo{volume}{\textbf{44}}},
  \bibinfo{pages}{1451} (\bibinfo{year}{1991}).

\bibitem[{\citenamefont{Nielsen et~al.}(2004)\citenamefont{Nielsen, Gartner,
  and Jahnke}}]{Nielsen:04}
\bibinfo{author}{\bibfnamefont{T.~R.} \bibnamefont{Nielsen}},
  \bibinfo{author}{\bibfnamefont{P.}~\bibnamefont{Gartner}}, \bibnamefont{and}
  \bibinfo{author}{\bibfnamefont{F.}~\bibnamefont{Jahnke}},
  \bibinfo{journal}{Phys. Rev. B} \textbf{\bibinfo{volume}{\textbf{69}}},
  \bibinfo{pages}{235314} (\bibinfo{year}{2004}).

\bibitem[{\citenamefont{Gartner et~al.}(1999)\citenamefont{Gartner, B{\'a}nyai,
  and Haug}}]{Gartner:99}
\bibinfo{author}{\bibfnamefont{P.}~\bibnamefont{Gartner}},
  \bibinfo{author}{\bibfnamefont{L.}~\bibnamefont{B{\'a}nyai}},
  \bibnamefont{and} \bibinfo{author}{\bibfnamefont{H.}~\bibnamefont{Haug}},
  \bibinfo{journal}{Phys. Rev. B} \textbf{\bibinfo{volume}{\textbf{60}}},
  \bibinfo{pages}{14234} (\bibinfo{year}{1999}).

\end{thebibliography}

\end{document}